\documentclass[useAMS,usenatbib,
onecolumn
]{mn2e}
\usepackage{graphicx}
\usepackage{amssymb,amsmath}
\usepackage{natbib}
\usepackage{bm}

\usepackage[breaklinks]{hyperref} 
\def\pFa{p_{\raise-0.3ex\hbox{{\scriptsize F$\!$\raise-0.03ex\hbox{$i$}}}}
}  
\def\pFas{p_{\raise-0.3ex\hbox{{\scriptsize F$\!$\raise-0.03ex\hbox{$j$}}}}
}  
\def\vFa{v_{\raise-0.3ex\hbox{{\scriptsize F$\!$\raise-0.03ex\hbox{$i$}}}}
}  
\newcommand{\mdens}{{\rm g~cm^{-3}}}
\newcommand{\bdens}{{\rm fm^{-3}}}
\newcommand{\msun}{{\rm M}_\odot}

\title[Physics input for modelling superfluid neutron stars with hyperon cores]
{Physics input for modelling superfluid neutron stars with hyperon cores}
\author[M. E. Gusakov et al.]
{M.~E.~Gusakov$^{1,2}$\thanks{gusakov@astro.ioffe.ru},
P.~Haensel$^3$\thanks{haensel@camk.edu.pl},
 E.~M.~Kantor$^{1}$\thanks{kantor@mail.ioffe.ru},\\
$^1$ Ioffe Physical-Technical Institute of the Russian Academy of
Sciences, Polytekhnicheskaya 26, 194021 Saint-Petersburg, Russia\\
$^2$ Saint-Petersburg State Polytechnical University,
Polytekhnicheskaya 29, 195251 St.-Petersburg, Russia\\
$^3$ N. Copernicus Astronomical Center, Polish Academy of Sciences, Bartycka 18,
PL-00-716 Warszawa, Poland}

\begin{document}

\date{Accepted 2013 xxxx. Received 2013 xxxx;
in original form 2013 xxxx}

\pagerange{\pageref{firstpage}--\pageref{lastpage}} \pubyear{2013}

\maketitle

\label{firstpage}

\begin{abstract}
Observations of massive ($M \approx 2.0~\msun$) neutron stars (NSs), 
PSRs J1614-2230 and J0348+0432,
rule out most of the models of nucleon-hyperon matter 
employed in NS simulations.
Here we construct three possible models of nucleon-hyperon matter consistent 
with the existence of $2~\msun$ pulsars as well as with 
semi-empirical nuclear matter parameters at saturation, and semi-empirical
hypernuclear data. 
Our aim is to calculate for these models 
all the parameters necessary for modelling dynamics of hyperon stars
(such as equation of state, adiabatic indices, thermodynamic derivatives, 
relativistic entrainment matrix, etc.),
making them available for a potential user.
To this aim a general non-linear hadronic Lagrangian involving 
$\sigma\omega\rho\phi\sigma^\ast$ meson fields,
as well as quartic terms in vector-meson fields, 
is considered. 
A universal scheme for calculation of the $\ell=0,1$ Landau Fermi-liquid parameters
and relativistic entrainment matrix 
is formulated in the mean-field approximation.
Use of this scheme allow us to obtain
numerical tables with 
the equation of state, Landau quasiparticle effective masses, 
adiabatic indices, the $\ell=0,1$ Landau Fermi-liquid parameters, 
and the relativistic entrainment matrix for the selected models of nucleon-hyperon matter.
These data are available on-line and suitable for numerical implementation 
in computer codes modelling various dynamical processes in NSs, 
in particular, oscillations of superfluid NSs and their cooling.
\end{abstract}


\maketitle

\section{Introduction}
\label{1}

Neutron stars (NSs),  being massive,
 compact, rapidly rotating objects,  with central density
up to ten times normal nuclear density ($\rho_{0}\approx 2.8\times 10^{14}~\mdens$,
corresponding to baryon number density $n_{0}\approx 0.16~\bdens$), are promising sources of
gravitational waves, associated with axial-symmetry breaking  stellar pulsations,
triggered by various types of instabilities (\citealt{andersson_et_al_11, andersson_et_al_13}). 
Modelling NS dynamics requires hydrodynamical description of its liquid core, of
density ranging from $\sim 0.5\rho_{0}$ at the outer edge of the core, to $\sim
10\rho_{0}$ at the centre of the most massive stars. It is expected that  the core
layer up to $2-3\rho_{0}$, called the outer core, consists of nucleons (mostly
neutrons) and leptons (electrons and muons), while at higher density (inner core)
the matter is expected to contain also hyperons. We are then dealing with a baryon
matter, consisting of more than two baryon species (to be contrasted with nuclear
matter in the outer core),  with an admixture of leptons required by
weak-interaction equilibrium and charge neutrality. At least some of the baryon
species are thought to be superfluid.

To study dynamics
of a multi-superfluid nucleon-hyperon (NH) matter, 
one needs not only the equation of state (EOS), involving various thermodynamic derivatives, 
but also a (symmetric) relativistic entrainment
matrix $Y_{ij}$ (hereafter subscripts $i, \, j$ run over all baryon species), 
describing non-dissipative interaction
between superfluids due to strong interaction of baryons. A method of the
calculation of $Y_{ij}$ for a mixture of NH superfluids was presented
in the limiting case of zero temperature ($T=0$) in \cite*{gkh09a} and then
generalized to non-zero $T$ in \cite*{gkh09b}. 
Strong interactions between
baryons were included using relativistic extension (\citealt{bc76}) 
of the Landau theory of Fermi liquids.

Numerical results of \cite{gkh09a,gkh09b} were obtained employing a basic version of
the relativistic mean field model 
(RMF; see \citealt{glendenning00, glendenning85} and references therein). 
This RMF model involved the baryon octet, interacting  via coupling to
scalar ($\sigma$), vector ($\omega^\mu$), and vector-isovector ($\rho^\mu_a$)
meson fields; here $\mu$ and $a$ indices denote the spacetime and isospin
components of the field, respectively. Unfortunately, the model used in \cite{gkh09a,gkh09b}
is not consistent with up-to-date hypernuclear data.

In the present paper we  replace the $\sigma\omega\rho$ Lagrangian by  a more
general non-linear model involving two additional hidden-strangeness mesons
(\citealt{bm09} and references therein). In this way we are able to overcome
the shortcomings of \cite{gkh09a,gkh09b}. Our models  fulfil constraint on the maximum NS mass 
$M_{\rm max}>2~\msun$ resulting from the discovery of   $2~\msun$ pulsars
(\citealt{demorest_et_al_10,antoniadis_et_al_13}). They are consistent with semi-empirical
saturation parameters of nuclear matters, binding energies of $\Lambda$ and $\Xi^-$ hyperons
in nuclear matter 
deduced from hypernuclei, and reproduce potential well for $\Sigma^-$ in nuclear matter
deduced from the $\Sigma^-$ atoms. 
As shown in several recent papers (\citealt{bhzbm12}; \citealt*{wcs12a,wcs12b}), 
all these constraints can be simultaneously satisfied by introducing an  additional vector
meson field $\phi$ coupled only to hyperons, resulting in  a strong  hyperon
repulsion at high densities, and/or allowing for breaking of SU(6) symmetry in the
vector-mesons coupling to hyperons. Therefore, instead of a
(too) simple  $\sigma\omega\rho$ model, used in \cite{gkh09a,gkh09b}, we consider
at least the $\sigma\omega\rho\phi$ one. 
In order to get a better fit to a larger
number of semi-empirical hyper-nuclear parameters 
(e.g., to describe a weak $\Lambda-\Lambda$ interaction 
following from `Nagara' event, see \citealt{takahashi_et_al_01}), 
an additional scalar meson $\sigma^\ast$ can be included, 
leading to a $\sigma\omega\rho\phi\sigma^\ast$
model. 
For the general
$\sigma\omega\rho\phi\sigma^\ast$ model Lagrangian that includes quartic terms in
vector-meson fields
we develop  a calculational scheme for the
$f_1^{ij}$ Landau Fermi-liquid parameters, and associated with them matrix $Y_{ij}$, 
as well as for the $f_0^{ij}$ Landau parameters
needed for calculation of various thermodynamic derivatives.
Numerical calculations of $f_0^{ij}$, $f_1^{ij}$ and $Y_{ij}$ are done for three selected
models of dense NH matter consistent 
with existence of $2~\msun$ pulsars as well as with 
semi-empirical nuclear matter parameters at saturation, and semi-empirical
hypernuclear data. 
For these models we also present EOS, Landau effective masses of baryons, and
adiabatic indices. 
This data provide all microphysics input allowing one
to model dynamics of superfluid NSs.
All numerical results are available on-line. 

The plan of this paper is as follows. Basic  definitions and relations for
superfluid NH mixture are recapitulated in Sect.\ \ref{sect:def.rel}. The
$\sigma\omega\rho\phi\sigma^\ast$ Lagrangian for the baryon octet is presented in
Sect.\ \ref{sect:lagrangian}. The Dirac equations for baryons and their solutions in
the RMF approximations are given in Sect.\ \ref{sect:baryons}. 
The equations for the
meson fields in the presence of baryon currents are  given in
Sect.\ \ref{sect:mesons}. 
Sect.\ \ref{sec:chempot} presents expressions for thermodynamic functions. 
Landau parameters 
$f_1^{ij}$ and $f_0^{ij}$ are derived in Sect.\ \ref{sect:f1} and  \ref{sect:f0},
respectively. Numerical results are collected in  Sect.\ \ref{sect:results}. 
Three up-to-date RMF models of NH matter are presented in Sect.\ \ref{sect:EOS.Mmax}. 
The EOSs for these models as well as the parameters
of NS configurations with maximum mass $M_{\rm max}$
are compared in the same Sect.\ \ref{sect:EOS.Mmax}.
Particle fractions for NH matter in beta equilibrium, 
adiabatic indices, and the speed of sound, all as functions of baryon number density, 
are compared in Sect.\ \ref{sect:yi.vsound}. 
Landau effective masses are calculated in Sect.\  \ref{sect:m.eff}. 
Numerical results for the Landau Fermi-liquid parameters 
and entrainment matrix are presented in Sect.\ \ref{sect:F0}. 
Stability of the ground state of NH matter is briefly discussed 
in Sect.\ \ref{sect:stab.F0.F1}.
Section \ref{sect:discussion} contains  
summary of our results.
Detailed information about the coupling constants 
for the three RMF models employed in this paper is given in
Appendix \ref{appendix:coupling}.
The way of calculating EOSs for these three models 
is reviewed in Appendix \ref{appendix:eos}.
Adiabatic indices are discussed in Appendix \ref{app:adiabat}.
Finally, a description of publicly available on-line numerical material 
containing the results of our calculations is given in Appendix \ref{appendix:online}.

\section{Basic definitions and relations}
\label{sect:def.rel}

Here we briefly review the Landau Fermi-liquid theory (see, e.g., \citealt{bp91, pn99})
generalized to the case of relativistic one-component liquid by \cite{bc76}
and extended to relativistic mixtures by \cite{gkh09a}.
For the sake of compactness of notation we use the convention $\hbar=c=1$, 
where $\hbar$ is the Planck constant and $c$ is the speed of light;
we also assume that the metric tensor is
$\eta_{\mu \nu}={\rm diag}(1, -1, -1, -1)$.
Unless otherwise stated, 
all quantities and relations are given 
in the reference system associated 
with normal fluid of leptons. 
In this reference system, 
the four-velocity of the normal fluid is $u^\mu=(1,0,0,0)$.

We are dealing with an uniform  mixture of  baryon species. 
Let us first assume that all baryon species are  {\it normal} (no superfluid gaps). This
means that if we start with a (reference) system of bare noninteracting baryons
and then the interaction is slowly switched
on, the system of bare noninteracting baryons transforms adiabatically into a
system of {\it Landau quasiparticles}. This system of quasiparticles retains essential properties
of a mixture of ideal Fermi gases. Namely the distribution function of
quasiparticles in the momentum space is the same as that of
an ideal `reference' system. The number of quasiparticles is equal to the
number of particles. The Landau Fermi-liquid
theory establishes therefore a one to one correspondence between the states of a
system of quasiparticles  and those of  the real system. The quasiparticle
states will be labelled by momentum $\pmb p$ and spin $s$, ${\pmb p}s$. As we will deal with spin
unpolarized systems, the quantities under consideration are spin independent and
can be replaced by spin-averaged ones.

In the ground
state, distribution function of quasiparticle species $i$ 
is then a filled Fermi sphere,
\begin{equation}
n_{i0}({\pmb p}) = \theta(\pFa - p), \label{Fermisphere}
\end{equation}
where $\pFa$ is the Fermi momentum
for quasiparticles, coinciding with that for bare noninteracting particles,
so that the number density $n_i=\pFa^3/3{\mathrm \pi}^2$.
$\theta(x)$ is the step function: $\theta(x) =1$, if $x>0$ and $0$ otherwise.
Subscript $0$ will denote the quantities in the ground state of the system.

Within the normal Landau Fermi-liquid theory, the energy of the system is a functional of
the quasiparticle distribution functions, $n_{i}({\pmb p})$. The validity of the
quasiparticle description of an excited state is restricted to
vicinity of the Fermi surfaces. This
means, that $\delta n_i({\pmb p})=n_{i}({\pmb p})-n_{i0}({\pmb p})$ is nonzero for
$\vert p-\pFa\vert\ll \pFa$. 
The energy $E$ of an excited state of
the system can be expressed in terms of $\delta n_i({\pmb p})$ by expanding the
functional $E\lbrace n_i({\pmb p})\rbrace $
around 
$E_0$
(see, e.g., \citealt{bp91, pn99}),
\begin{equation}
E-E_{0} = \sum_{{\pmb p} s i} \varepsilon_{i0}({\pmb p}) \, \delta n_i({\pmb p}) +
\frac{1}{2} \sum_{{\pmb p} {\pmb p}' \, ss' \, ij} f^{ij}({\pmb p}, {\pmb p}') \,
\delta n_i({\pmb p}) \delta n_j({\pmb p}')~,
\label{energy2}
\end{equation}
where third order terms in $\delta n_i({\pmb p})$ have been neglected. 
Here,
$\varepsilon_{i0}({\pmb p})$
and $f^{ij}({\pmb p}, {\pmb p}')$
are, respectively, energy of a $i$-quasiparticle in the ground state, 
and 
the (spin-averaged) quasiparticle interaction -- a central object in the 
Landau Fermi-liquid theory. 
Because of the isotropy of the ground state, $\varepsilon_{i0}({\pmb p})$ depends
only on $\vert{\pmb p}\vert=p$. 
This needs not to be so for a quasiparticle
in an excited state of the system; in the latter case 
the quasiparticle energy is given by (\citealt{bp91, pn99})
\begin{equation}
\varepsilon_{i}({\pmb p})
= \varepsilon_{i0}(p)+ \sum_{{\pmb p}' s' j} f^{ij}({\pmb p}, {\pmb p}') \,
\delta n_j({\pmb p}')~.
\label{eq:e_i}
\end{equation}

Near the Fermi surface, 
the function
$\varepsilon_{i0}(p)$ can be expanded into a
series in powers of the quantity $p-\pFa$ and
approximated by a linear form,
\begin{equation}
\varepsilon_{i0}(p) \approx \mu_i+ \vFa (p-\pFa), \label{energy0}
\end{equation}
where  $\mu_i= \varepsilon_{i0}(\pFa)$ is the relativistic (i.e., including the
rest energy) chemical potential or, equivalently, the Fermi energy of quasiparticle
species $i$ and 
$v_{{\rm F}i} = [\partial \varepsilon_{i0}(p)/\partial p]_{p=\pFa}$ 
is the velocity of quasiparticles on the Fermi
surface.  The Landau quasiparticle effective mass $m^{\ast}_i$ is introduced
through the relation
\begin{equation}
\vFa \equiv \pFa/m^{\ast}_i~.
\label{eq:m^ast}
\end{equation}

Within the region of validity of the Landau Fermi-liquid theory, the magnitude
of momentum arguments  of the quasiparticle interaction $f^{i j}({\pmb p},
{\pmb p}')$ can be approximated as  $\vert {\pmb p}\vert\approx \pFa$ and
$\vert{\pmb p}^\prime\vert \approx \pFas$, respectively. Therefore, the momentum dependence of  $f^{i
j}$ can be expanded into Legendre polynomials $P_\ell({\rm cos \, \theta})$,
\begin{equation}
f^{i j}({\pmb p}, {\pmb p}') = \sum_\ell f^{i j}_\ell \, P_\ell(\cos \theta),
\label{fik}
\end{equation}
where $\theta$ is the angle between ${\pmb p}$ and ${\pmb p}'$ and $f^{i j}_\ell$
are the Landau Fermi-liquid parameters, $f_\ell^{ij}=f_\ell^{ji}$. The
dimensionless Landau parameters $F^{ij}_\ell$ are defined through
\begin{equation}
F^{ij}_\ell\equiv  \sqrt{N_{{\rm F}i}N_{{\rm F}j}}f^{ij}_\ell~,
~ N_{{\rm F}i}= m^\ast_i p_{{\rm F}i}/{\mathrm \pi}^2~,
\label{eq:F.f}
\end{equation}
where $N_{{\rm F}i}$ is the density of $i$-quasiparticle states at the
Fermi surface.

The effective mass $m_i^{\ast}$ in the
relativistic theory can be expressed in terms of the Landau parameters
$F^{i j}_1$ [see equation (24) of \citealt{gkh09a}],
\begin{equation}
\frac{\mu_i}{m_i^{\ast}} = 1 -
\frac{1}{3}
\sum_j \frac{\mu_j}{\sqrt{m_i^\ast m_j^\ast}}
\left(\frac{p_{{\rm F}j}}{p_{{\rm F}i}}\right)^{3/2}F^{ij}_1 .
\label{effmass}
\end{equation}

Let us pass now to the {\it superfluid baryons}. A mixture of
baryon superfluids is  described in  terms of {\it Bogoliubov quasiparticles}.
At $T=0$ all baryons are paired into Cooper pairs
and the  energy gaps for (Bogoliubov) $i$-quasiparticles at the Fermi surface
are $\Delta_i$ (for the sake of simplicity we restrict ourselves
to isotropic  gaps). As long as $\Delta_i\ll \mu_i-m_i$, 
the energy gaps will not affect the main formulas,
e.g., the  particle current
densities  ${\pmb j}_i$ will be related to the 
distribution functions by the same expression as in the case of a normal
Fermi liquid, see \cite{leggett65,leggett75}.
 
We  consider  excited states of the system  associated
with  uniform superfluid flows, each of them  with macroscopic
velocity ${\pmb V}_{{\rm s}i}$. The macroscopic flow velocity of species
$i$ is related to the total momentum of a Cooper pair $2 \, {\pmb Q}_i$ by
\begin{equation}
{\pmb V}_{{\rm s}i} = \frac{{\pmb Q}_i}{m_i},
\label{VsQ}
\end{equation}
where $m_i$ is a  free  (in vacuum) mass of baryon $i$.
In the linear approximation in ${\pmb Q}_i$ ($\vert {\pmb Q}_i\vert \ll p_{{\rm F}i}$), the current
densities  ${\pmb j}_i$  are connected with  $\lbrace {\pmb Q}_j \rbrace$
by (\citealt{gkh09a})
\begin{equation}
{\pmb j}_i = \sum_j \, Y_{ij} \, {\pmb Q}_j.
\label{jnn}
\end{equation}
The relativistic  entrainment matrix, $Y_{ij}$,
is symmetric, $Y_{ij}=Y_{ji}$, and fulfils a sum rule
\begin{equation}
\sum_j \mu_j \, Y_{ij} = n_i~.
\label{sumrule}
\end{equation}
In the case of vanishing temperature ($T=0$),
the entrainment matrix can be expressed in terms of the $F^{ij}_1$ Landau
parameters (\citealt{gkh09a}),
\begin{equation}
Y_{ij}=\frac{n_i}{m_i^\ast}\delta_{ij}+
\frac{1}{3}\left(\frac{n_i n_j}{m_i^\ast m_j^\ast}\right)^{1/2}
F^{ij}_1~,
\label{eq:Y.F}
\end{equation}
where $\delta_{ij}$ is the Kronecker delta.
The non-diagonal elements of $Y_{ij}$ describe the superfluid entrainment, 
a non-dissipative interaction between the superfluid baryon flows.

The more complex expression for $Y_{ij}$, valid at arbitrary temperature, 
was formulated by \cite{gkh09b}. 
It also involves the Landau parameters $F_1^{ij}$
(see Eqs.\ (41)--(43) and (45) of that reference).
 
\section{The \texorpdfstring{$\sigma\omega\rho\phi\sigma^\ast$}{Lg}  model}
\label{sect:RMF.model}

We use  a nonlinear model of \cite{bm09}.

\subsection{Lagrangian}
\label{sect:lagrangian}

The Lagrangian density ${\cal L}$ of the strongly interacting baryon
 system can be split into two basic components,
${\cal L}_{\rm BM}$ involving baryon terms  affected by the meson fields, and
${\cal L}_{\rm M}$ involving exclusively meson fields.

We consider Dirac spinor fields for baryons  $\Psi_i$,
depending on spacetime point $x$. The contravariant coordinates of $x$ are
$x^\mu=(x^0,x^1,x^2,x^3,x^4)$, and the contravariant derivative $\partial^\mu=
\partial/\partial x_\mu$. The ${\cal L}_{\rm BM}$ component of the Lagrangian density
is then
\begin{equation}
{\cal L}_{\rm BM}=\sum_i\overline{\Psi}_i\left({\rm i}\gamma_\mu \partial^\mu
-m_i
-g_{\omega i}\gamma_\mu\omega^\mu -
g_{\phi i}\gamma_\mu\phi^\mu-
 \frac{1}{2}g_{\rho i}\tau_a\gamma_\mu\rho_a^\mu
+g_{\sigma i}\sigma
+g_{\sigma^\ast i}\sigma^\ast\right)\Psi_i~,
\label{eq:L.BM}
\end{equation}
where $\gamma_\mu$ are Dirac matrices and $\overline{\Psi}_i\equiv
\Psi^\dagger_i\gamma_0$ is an
adjoint Dirac spinor. 
The components of $\tau_a$ $(a=1,2,3)$ are Pauli
matrices acting in the isospin space. 
The parameters $g_{mi}$ ($m=\sigma$, $\sigma^\ast$, $\omega$, $\rho$, $\phi$) 
are coupling constants
of the meson fields to the baryon fields.

Baryon densities and currents are sources of the meson fields. We assume
spatially uniform and time independent sources and therefore resulting meson fields are
$x$-independent. Meson fields and
their interactions generate the meson Lagrangian 
${\cal L}_{\rm M}={\cal L}_{\rm M}^{\rm(S)}+{\cal L}_{\rm M}^{\rm (V)}$, 
where S and V
refer to the scalar and vector
mesons, respectively. The scalar meson contribution is
\begin{equation}
{\cal L}_{\rm M}^{\rm (S)}=
-\frac{1}{2}m_{\sigma}^{2}\sigma^{2}-\frac{1}{2}m_{\sigma^{\ast}}^{2}
\sigma^{\ast2}-
 \frac{1}{3}g_{3}\sigma^{3}-\frac{1}{4}g_{4}\sigma^{4},
 \label{eq:L.M.S}
\end{equation}
where the coupling constants $g_3$ and $g_4$ determine
the strength of $\sigma$ meson self-interaction.

The vector meson contribution includes terms quadratic and quartic in vector meson
fields,
\begin{eqnarray}
{\cal L}_{\rm M}^{\rm (V)}&=&
\underline{\frac{1}{2}m_{\omega}^{2}(\omega_{\mu}\omega^{\mu})}
 + \underline{\frac{1}{2}m_{\rho}^{2}(\rho_{a \mu}\rho^{\mu}_a)}
 +\underline{\frac{1}{2}m_{\phi}^{2}(\phi_{\mu}\phi^{\mu})}
 +\underline{\frac{1}{4}c_{3}(\omega_{\mu}\omega^{\mu})^{2}}
 \nonumber\\
&&+\frac{1}{4}\widetilde{c}_{3}(\rho_{a \mu}\rho^{\mu}_a)^{2}
+\frac{3}{4}\widetilde{c}_{3}(\rho_{a \mu}\rho^{\mu}_a)(\phi_{\nu}\phi^{\nu})
+\frac{1}{8}\widetilde{c}_{3}(\phi_{\mu}\phi^{\mu})^{2} 
\nonumber \\ 
&&+\frac{3}{4}\widetilde{c}_{3}(\phi_{\mu}\phi^{\mu})(\omega_{\nu}\omega^{\nu})
+\frac{1}{4}(g_{\rho {\rm N}}g_{\omega {\rm N}})^{2}\Lambda_{\rm V} (\phi_{\mu}\phi^{\mu})^{2}
-\frac{1}{2}(g_{\rho {\rm N}}g_{\omega {\rm N}})^{2}\Lambda_{\rm V}
 (\rho_{a \mu}\rho^{\mu}_a)
 (\phi_{\nu}\phi^{\nu}) 
 \nonumber \\
  & & +(g_{\rho {\rm N}}g_{\omega {\rm N}})^{2}\Lambda_{\rm V}(\rho_{a \mu}\rho^{\mu}_a)
 (\omega_{\nu}\omega^{\nu})-\frac{1}{2}(g_{\rho {\rm N}}g_{\omega {\rm N}})^{2}
 \Lambda_{\rm V}(\phi_{\mu}\phi^{\mu})(\omega_{\nu}\omega^{\nu}).
 \label{LMV}
 \end{eqnarray}
In Eqs.\ (\ref{eq:L.M.S}) and (\ref{LMV})
$m_\sigma$, $m_{\sigma^\ast}$, $m_{\omega}$, $m_{\rho}$, and $m_{\phi}$
are the corresponding meson masses.
Three additional parameters in Eq. (\ref{LMV}), 
$c_3$, $\widetilde{c}_3$,  and $\Lambda_{\rm V}$, 
determine the strength of the quartic vector meson terms.
(Notice that in \citealt{bm09} 
it was assumed that $\widetilde{c}_3=c_3$.)
For less general models considered by us in Sect.\ \ref{sect:results}
only the underlined terms in Eq.\ (\ref{LMV}) are taken 
into account, that is we put $\widetilde{c}_3=\Lambda_{\rm V}=0$.

\subsection{The microscopic state of baryons in the RMF approximation}
\label{sect:baryons}

The equations of motion for the baryon fields $\Psi_i$ are obtained as
Euler-Lagrange equations from ${\cal L}$,
\begin{equation}
\partial {\cal L}/\partial \overline{\Psi}_i=\left({\rm i}\gamma_\mu \partial^\mu -m_i
-g_{\omega i}\gamma_\mu \omega^\mu -
g_{\phi i}\gamma_\mu\phi^\mu-
\frac{1}{2} g_{\rho i}\tau_a\gamma_\mu\rho_a^\mu
+g_{\sigma i}\sigma+
g_{\sigma^\ast i}\sigma^\ast\right)\Psi_i=0~.
\label{eq:EL.Psi_j}
\end{equation}
These are Dirac equations for baryons coupled to  meson fields.
We look for  the macroscopic states of NH matter which are uniform in space and
stationary. In the RMF approximation, the meson fields in ${\cal L}$
are replaced by their  $x$-independent mean values. Therefore,
solutions $\Psi_i$ of Eq.\;(\ref{eq:EL.Psi_j}) are the
eigenstates of the four-momentum $p^\mu$,
\begin{equation}
\Psi_i=\Psi_i(p^\mu){\rm  e}^{-{\rm i}p^\mu x_\mu}~.
\label{eq:Psi_j}
\end{equation}

After putting Ansatz (\ref{eq:Psi_j}) into the equation of motion
(\ref{eq:EL.Psi_j}), we solve it using standard methods for the Dirac equation. In
this way we find the {\it Dirac equation eigenvalues} of the  energy, $e_i$, at
fixed values of the uniform meson fields,
\begin{eqnarray}
e_{i}(\pmb{p}) &=& g_{\omega i}\omega^0+
g_{\rho i}I_{3i}\rho^0_3+g_{\phi i}\phi^0 
\nonumber\\
&+& \left[\left(\pmb{p}-g_{\omega i}\pmb{\omega}-
g_{\rho i}I_{3i}\pmb{\rho}_3-g_{\phi i}\pmb{\phi}\right)^2
+ (m_i - g_{\sigma i}\sigma
-g_{\sigma^\ast i}\sigma^\ast)^2 \right]^{1/2},
\label{eq:epsilon.i}
\end{eqnarray}
where $I_{3i}$ is the third component
of the isospin of baryon $i$, with $I_{3{\rm n}}=-1/2$ 
(the subscript n stands for neutrons).

A macroscopic spatially uniform stationary state for baryons, under given
constraints on baryon currents, $\pmb{j}_i$, and baryon densities, $n_i$, is
obtained by filling lowest Dirac energy eigenstates. The distribution function of
the occupied Dirac states coincides then with distribution function of the Landau
$i$-quasiparticles. Therefore, in the RMF approximation, the quasiparticle energy
of a baryon species $i$ is equal to the Dirac equation eigenvalue,
$\varepsilon_i({\pmb p})=e_i({\pmb p})$.
In particular, the particle current density can be expressed through
$e_{i}({\pmb p})$ as
\begin{equation}
{\pmb j}_i = \sum_{{\pmb p} s} \frac{\partial e_{i}(\pmb p)}{\partial {\pmb p}} \, n_i({\pmb p}).
\label{j}
\end{equation}
%

\subsection{Field equations for meson fields in the presence of baryon currents}
\label{sect:mesons}

Meson fields are calculated assuming a uniform stationary state of baryons.
The field equations for mesons are the Euler-Lagrange equations obtained
from ${\cal L}$. The baryon fields
enter the source terms in the meson field equation. 
In the RMF approximation, 
the source term is replaced by a mean value calculated in the
uniform stationary state of the baryon system described in
Sect.\ \ref{sect:baryons}.
Both source terms and meson fields are $x$-independent.
Equations for meson fields can be written as 
\begin{eqnarray}
&&m_\sigma^2 \, \sigma =
-g_3 \, \sigma^2 - g_4 \, \sigma^3 + \sum_{i}
g_{\sigma i} \,
R_i(m_i-g_{\sigma i} \, \sigma - g_{\sigma^* i} \, \sigma^*, \,\,
g_{\omega i} \, {\pmb \omega} +g_{\rho i}\, I_{3i} \, {\pmb \rho}_3 + g_{\phi i }\, {\pmb \phi}),
\label{sigma} \\
&&m_{\sigma^*}^2 \, \sigma^* = \sum_i g_{\sigma^* i} \,
R_i(m_i-g_{\sigma i} \, \sigma - g_{\sigma^* i} \, \sigma^*, \,\,
g_{\omega i} \, {\pmb \omega} +g_{\rho i}\, I_{3i} \, {\pmb \rho}_3+g_{\phi i }\, {\pmb \phi}),
\label{sigma*}\\
&& \left[ m_\omega^2 + A_\omega \, (\omega_\nu \omega^\nu) + A_{\rho} \, (\rho_{3 \, \nu} \rho_3^{\nu})
+A_{\phi} \, (\phi_\nu \phi^\nu)
\right] \, \omega^{\mu} = \sum_i g_{\omega i} \, j_i^{\mu},
\label{omega}\\
&& \left[ m_\rho^2 + B_\omega \, (\omega_\nu \omega^\nu) + B_{\rho} \, (\rho_{3 \, \nu} \rho_3^{\nu})
+B_{\phi} \, (\phi_\nu \phi^\nu)
\right] \, \rho_3^{\mu} = \sum_i g_{\rho i} \, I_{3i} \, j_i^{\mu},
\label{rho}\\
&& \left[ m_\phi^2 + C_\omega \, (\omega_\nu \omega^\nu) + C_{\rho} \, (\rho_{3 \, \nu} \rho_3^{\nu})
+C_{\phi} \, (\phi_\nu \phi^\nu)
\right] \, \phi^{\mu} = \sum_i g_{\phi i} \, j_i^{\mu}.
\label{phi}
\end{eqnarray}
Here
\begin{equation}
R_i(x, {\pmb y}) = \sum_{{\pmb p} s} \frac{x}{\sqrt{({\pmb p}-{\pmb y})^2 + x^2}} \, n_i({\pmb p}).
\label{R}
\end{equation}
In case of the $\sigma\omega\rho\phi\sigma^\ast$ Lagrangian of \cite{bm09},
described in Sect.\ \ref{sect:lagrangian},
the constants $A_\omega, \ldots, C_{\phi}$
in Eqs.\ (\ref{omega})--(\ref{phi}) are given by expressions
\begin{eqnarray}
&& A_\omega = c_3, \quad
A_{\rho} = 2 \, \Lambda_{\rm V} \, (g_{\omega {\rm N}} g_{\rho {\rm N}})^2, \quad
A_{\phi} = \frac{3}{2} \widetilde{c}_3 - \Lambda_{\rm V} \,(g_{\omega {\rm N}} g_{\rho {\rm N}})^2,
\label{A}\\
&&B_\omega = 2 \, \Lambda_{\rm V} \, (g_{\omega {\rm N}} g_{\rho {\rm N}})^2, \quad
B_{\rho} = \widetilde{c}_3, \quad
B_{\phi} =A_{\phi},
\label{B}\\
&& C_{\omega} = \frac{3}{2} \widetilde{c}_3 - \Lambda_{\rm V} \,(g_{\omega {\rm N}} g_{\rho {\rm N}})^2, \quad
C_{\rho} = C_{\omega}, \quad
C_{\phi} = \frac{1}{2} \widetilde{c}_3 + \Lambda_{\rm V} \,(g_{\omega {\rm N}} g_{\rho {\rm N}})^2.
\label{C}
\end{eqnarray}
As we already emphasized above, 
for less general RMF models considered in Sect.\ \ref{sect:results}, 
the only non-zero constant is $A_{\omega}=c_3$ 
($\widetilde{c}_3=\Lambda_{\rm V}=0$).
Neglecting $\sigma^\ast$ and $\phi$ mesons, 
Eqs.\ (\ref{sigma})--(\ref{C}) 
correctly reproduce the $\sigma\omega\rho$ model of Glendenning 
(\citealt{glendenning00, gkh09a}).

\subsection{Chemical potential, energy density, and pressure}
\label{sec:chempot}

In this section, we assume that there is no baryon currents in the system,
that is, ${\pmb w}={\pmb \rho_3}={\pmb \phi}=0$.
Knowledge of the particle energy (\ref{eq:epsilon.i}) 
allows one to immediately find 
the relativistic chemical potential $\mu_i$,
\begin{equation}
\mu_i=e_i(p_{{\rm F} i})=g_{\omega i}\omega^0+
g_{\rho i}I_{3i}\rho^0_3+g_{\phi i}\phi^0 +
\sqrt{p_{{\rm F} i}^2
+ (m_i - g_{\sigma i}\sigma
-g_{\sigma^\ast i}\sigma^\ast)^2}
\label{mui}
\end{equation}
and the Landau effective mass $m^\ast_i$ 
(cf. equation (47) of \citealt{gkh09a}),
\begin{equation}
m^\ast_i = \frac{p_{{\rm F }i}}{|\partial e_i(\pmb p)/\partial {\pmb p}|_{p=p_{{\rm F}i}}}
=\sqrt{p_{{\rm F}i}^2+(m_i-g_{\sigma i} \sigma -g_{\sigma^\ast i} \sigma^\ast)^2}. 
\label{meffLandau}
\end{equation}

The energy density $\rho$ (also termed {\it density} in what follows) 
can be obtained  from the Lagrangian of Sect.\ \ref{sect:lagrangian}
in the same way as it was done, e.g., in  \cite{glendenning00}.
The result is
\begin{eqnarray}
\rho=&-&\left\langle {\cal L} \right\rangle 
+ \sum_i 
\left[g_{\omega i} \, \omega^0 \, n_i
+ g_{\rho i} I_{3i} \, \rho_3^0 \, n_i
+ g_{\phi i} \, \phi^0 \, n_i
+ R_{\rm E}(m_i-g_{\sigma i} \, \sigma - g_{\sigma^\ast i} \, \sigma^\ast, \,\,p_{{\rm F}i})
\right]
\nonumber\\
&+& \sum_{l={\rm e}, \, \mu} R_{\rm E}(m_l, \,\,p_{{\rm F}l}),
\label{energy}
\end{eqnarray}
where the summation is performed over all baryon species $i$ and lepton species $l={\rm e}$, $\mu$;
\begin{eqnarray}
&&\left\langle {\cal L}\right\rangle =
-\frac{1}{2} m_{\sigma}^2 \sigma^2
- \frac{1}{3} g_3 \sigma^3
-\frac{1}{4} g_4 \sigma^4
+ \frac{1}{2} m_{\omega}^2 (\omega^0)^2
+ \frac{1}{2} m_{\rho}^2 (\rho_3^0)^2
+ \frac{1}{2} m_{\phi}^2 (\phi^0)^2
- \frac{1}{2} m_{\sigma^\ast}^2 \sigma^{\ast 2}
\nonumber\\
&&+ \frac{1}{4} c_3 \, (\omega^0)^4
+\Lambda_{\rm V} \, (g_{\omega {\rm N}} g_{\rho {\rm N}})^2 \, (\omega^0)^2 (\rho_3^0)^2
-\left[\frac{1}{2} \Lambda_{\rm V} \, (g_{\omega {\rm N}} g_{\rho {\rm N}})^2 -\frac{3}{4} \widetilde{c}_3 \right] 
\, (\omega^0)^2 (\phi^0)^2
\nonumber\\
&&-\left[\frac{1}{2} \Lambda_{\rm V} \, (g_{\omega {\rm N}} g_{\rho {\rm N}})^2 -\frac{3}{4} \widetilde{c}_3 \right] 
\, (\rho_3^0)^2 (\phi^0)^2
+\frac{1}{4} \widetilde{c}_3 (\rho_3^0)^4
+\left[\frac{1}{4} \Lambda_{\rm V} \, (g_{\omega {\rm N}} g_{\rho {\rm N}})^2 +\frac{1}{8} \widetilde{c}_3 \right] (\phi^0)^4,
\label{Lav}
\end{eqnarray}
and
\begin{equation}
R_{\rm E}(x,\, y)= \frac{1}{{\mathrm \pi}^2} \int_{0}^y p^2 \, \sqrt{p^2+x^2} \, {\rm d}p.
\label{RE}
\end{equation}
Now the pressure $P$ can be expressed through $\rho$ and $\mu_k$
by the following standard formula,
\begin{equation}
P=-\rho+\sum_k \mu_k n_k,
\label{pres}
\end{equation}
where the subscript $k$ runs over all particle species (baryons and leptons).

\section{Derivation of expression for \texorpdfstring{\lowercase{$f_1^{ij}$}}{Lg}  }  
\label{sect:f1}

To calculate the Landau parameter $f_1^{ij}$ we have to create a uniform  baryon
current in the system. For that we shift the distribution function $n_{i0}({\pmb
p})$ (the step function) of a baryon species $i$ by a small vector ${\pmb Q_i}$.
Note that, in the linear approximation in ${\pmb Q}_i$, the scalars $\sigma$,
$\sigma^*$, $\omega_\mu \omega^{\mu}$, $\rho_{3 \, \mu} \rho_3^{\mu}$, and
$\phi_{\mu} \phi^{\mu}$ remain the same as in the absence of baryon currents.

Following the derivation of equation (43) of  \cite{gkh09a}, one obtains
\begin{equation}
{\pmb j}_i = \frac{n_i}{m_i^*}
\, \left( {\pmb Q}_i - g_{\omega i} \, {\pmb \omega}
- g_{\rho i} \, I_{3i} \, {\pmb \rho}_3 - g_{\phi i} \, {\pmb \phi} \right),
\label{j2}
\end{equation}
where $m^\ast_i$ is given by Eq.\ (\ref{meffLandau}).
Eq.\ (\ref{j2}) should be supplemented by the expressions for ${\pmb \omega}$,
${\pmb \rho}_3$, and ${\pmb \phi}$.
These expressions can be found from Eqs.\ (\ref{omega})--(\ref{phi}),
\begin{eqnarray}
m_{\omega}^{*2} \, {\pmb \omega} &=& \sum_i g_{\omega i} \,\, {\pmb j}_i,
\label{omega2}\\
m_{\rho}^{*2} \, {\pmb \rho}_3 &=& \sum_i g_{\rho i} \, I_{3i} \,\, {\pmb j}_i,
\label{rho2}\\
m_{\phi}^{*2} \,{\pmb \phi} &=& \sum_i g_{\phi i} \,\, {\pmb j}_i,
\label{phi2}
\end{eqnarray}
where we defined the effective meson masses
\begin{eqnarray}
m^{*2}_\omega &=& m_\omega^2 + A_\omega \, (\omega^0)^2 + A_{\rho} \, (\rho_3^0)^2
+A_{\phi} \, (\phi^0)^2,
\label{momega}\\
m^{*2}_\rho &=& m_\rho^2 + B_\omega \, (\omega^0)^2 + B_{\rho} \, (\rho_3^0)^2
+B_{\phi} \, (\phi^0)^2,
\label{mrho}\\
m^{*2}_\phi &=& m_\phi^2 + C_\omega \, (\omega^0)^2 + C_{\rho} \, (\rho_3^0)^2
+C_{\phi} \, (\phi^0)^2,
\label{mphi}
\end{eqnarray}
and made use of the fact that, for example,
$\omega_{\mu} \omega^{\mu} = (\omega^0)^2$
with the accuracy to linear terms in ${\pmb Q}_i$
($\omega^0$ is also independent of ${\pmb Q}_i$ in the linear approximation).

Eq.\ (\ref{j2}) should be solved together with Eqs.\ (\ref{omega2})--(\ref{phi2}).
To proceed further, let us multiply (\ref{j2}) by $g_{\omega i}$
and sum it over $i$.
Then, using Eq.\ (\ref{omega2}), one obtains
\begin{equation}
m_\omega^{\ast 2}\, {\pmb \omega} = \sum_i \frac{n_i}{m_i^\ast} \,
g_{\omega i} \,
\left( {\pmb Q}_i - g_{\omega i} \, {\pmb \omega}
- g_{\rho i} \, I_{3i} \, {\pmb \rho}_3 - g_{\phi i} \, {\pmb \phi} \right).
\label{omega3}
\end{equation}
Similarly,
\begin{eqnarray}
m_\rho^{\ast 2} \, {\pmb \rho}_3 &=& \sum_i \frac{n_i}{m_i^\ast} \, g_{\rho i} \, I_{3i} \,
\left( {\pmb Q}_i - g_{\omega i} \, {\pmb \omega}
- g_{\rho i} \, I_{3i} \, {\pmb \rho}_3 - g_{\phi i} \, {\pmb \phi} \right),
\label{rho3}\\
m_\phi^{\ast 2} \, {\pmb \phi} &=& \sum_i \frac{n_i}{m_i^\ast} \, g_{\phi i} \,
\left( {\pmb Q}_i - g_{\omega i} \, {\pmb \omega}
- g_{\rho i} \, I_{3i} \, {\pmb \rho}_3 - g_{\phi i} \, {\pmb \phi} \right).
\label{phi3}
\end{eqnarray}
Eqs.\ (\ref{omega3})--(\ref{phi3}) can be rewritten in the matrix form,
\begin{equation}
\left(
\begin{array}{ccc}
m_{\omega}^{\ast 2}+\sum_i \frac{n_i}{m_i^\ast} \, g_{\omega i}^2 &
 \sum_i \frac{n_i}{m_i^\ast} \, g_{\omega i} \, g_{\rho i} \, I_{3i} &
 \sum_i \frac{n_i}{m_i^\ast} \, g_{\omega i} \, g_{\phi i}\\
\sum_i \frac{n_i}{m_i^\ast} \, g_{\omega i} \, g_{\rho i} \, I_{3i}&
m_{\rho}^{\ast 2}+\sum_i \frac{n_i}{m_i^\ast} \, g_{\rho i}^2 \, I_{3i}^2 &
\sum_i \frac{n_i}{m_i^\ast} \, g_{\rho i} \, I_{3i} \, g_{\phi i}\\
\sum_i \frac{n_i}{m_i^\ast} \, g_{\omega i} \, g_{\phi i} &
\sum_i \frac{n_i}{m_i^\ast} \, g_{\rho i} \, I_{3i}\,  g_{\phi i}&
m_{\phi}^{\ast 2}+ \sum_i \frac{n_i}{m_i^\ast} \, g_{\phi i}^2
\end{array}
\right)
\left(
\begin{array}{ccc}
{\pmb \omega} \\
 {\pmb \rho}_3 \\
{\pmb \phi}
\end{array}
\right) =
\left(
\begin{array}{ccc}
 \sum_i \frac{n_i}{m_i^\ast} \, g_{\omega i} \, {\pmb Q}_i \\
 \sum_i \frac{n_i}{m_i^\ast} \, g_{\rho i}\, I_{3i} \, {\pmb Q}_i\\
\sum_i \frac{n_i}{m_i^\ast} \, g_{\phi i} \, {\pmb Q}_i
\end{array}
\right).
 \label{matrix}
\end{equation}
The solution to Eq.\ (\ref{matrix}) can be presented as
\begin{eqnarray}
{\pmb \omega} &=& \sum_j \alpha_{\omega j} \, {\pmb Q}_{j},
\label{omega4}\\
{\pmb \rho}_3 &=& \sum_j  \alpha_{\rho j} \, {\pmb Q}_{j},
\label{rho4}\\
{\pmb \phi} &=& \sum_j  \alpha_{\phi j} \, {\pmb Q}_{j},
\label{phi4}
\end{eqnarray}
where the coefficients $\alpha_{\omega j}$, $\alpha_{\rho j}$, and $\alpha_{\phi j}$
can be determined from Eq.\ (\ref{matrix})
using the methods of linear algebra. 
Using Eqs.\ (\ref{j2}) and (\ref{omega4})--(\ref{phi4}),
one can calculate the entrainment matrix $Y_{ij}$,
\begin{equation}
Y_{ij} = \frac{n_i}{m_i^\ast}
\left(
\delta_{ij} - g_{\omega i} \, \alpha_{\omega j} - g_{\rho i} \, I_{3i} \, \alpha_{\rho j}
- g_{\phi i} \, \alpha_{\phi j} \right),
\label{Yij}
\end{equation}
and, consequently, the Landau parameters $f_1^{ij}$
[see equation (46) of \citealt{gkh09a} or Eq.\ (\ref{eq:Y.F})].

\section{Derivation of expression for \texorpdfstring{\lowercase{$f_0^{ij}$}}{Lg} } 
\label{sect:f0}

Here we will closely follow the derivation of section IIIC in \cite{gkh09a}.
Let us consider a system without baryon currents (i.e., ${\pmb \omega}={\pmb
\rho}_3={\pmb \phi}=0$). 
To calculate the Landau parameters $f_0^{ij}$ 
we slightly vary the Fermi momentum $p_{{\rm F}i}$ 
by a small quantity $\Delta p_{{\rm F}i}$,
so that the distribution function of a quasiparticle species $i$ will become 
$n_i({\pmb p})=\theta(p_{{\rm F} i} + \Delta p_{{\rm F}i}-p)$.
This will shift the energy of baryons $e_{i}({\pmb p})$ by a small quantity 
$\delta e_{i}({\pmb p})$.
At $p=p_{{\rm F}i}$ the expression for
$\delta e_{i}({\pmb p})$ takes the form [see Eq.\ (\ref{eq:epsilon.i})]
\begin{eqnarray}
\delta e_i(p_{{\rm F}i}) &=& g_{\omega i} \, \delta\omega^0
+ g_{\rho i} \, I_{3i} \, \delta\rho_3^0 + g_{\phi i} \, \delta \phi^0
\nonumber\\
&-& {(m_i/m_i^\ast-g_{\sigma i} \, \sigma/m_i^\ast -
g_{\sigma^* i} \, \sigma^*/m_i^\ast)}\,
(g_{\sigma i} \, \delta \sigma + g_{\sigma^\ast i} \, \delta \sigma^{\ast}),
\label{Ei}
\end{eqnarray}
where we used Eq.\ (\ref{meffLandau})
for the Landau effective mass $m_i^\ast$.
On the other hand, it follows from the Landau theory of Fermi liquids that
\begin{equation}
\delta e_i(p_{{\rm F}i}) = \sum_j f_0^{ij} \, \delta n_j,
\label{EiLandau}
\end{equation}
where $\delta n_j \equiv p_{{\rm F}j}^2 \Delta p_{{\rm F}j}/{\mathrm \pi}^2$.
Comparing Eqs.\ (\ref{Ei}) and (\ref{EiLandau}) one can calculate
the parameters $f_0^{ij}$.
For that we need to express
the variations $\delta \sigma$, $\delta \sigma^\ast$,
$\delta \omega^0$, $\delta \rho_3^0$, and $\delta \phi^0$
through $\delta n_i$.
We start with the quantities $\delta \sigma$ and $\delta \sigma^\ast$.
They can be found from the linearized Eqs.\ (\ref{sigma}) and (\ref{sigma*}),
\begin{eqnarray}
(m_{\sigma}^2+ 2 \, g_3 \, \sigma + 3 \, g_4 \, \sigma^2) \, \delta \sigma
= -\sum_i g_{\sigma i} \, \frac{\partial R_i(x, 0)}{\partial x}
|_{x=m_i-g_{\sigma i} \, \sigma - g_{\sigma^\ast i} \, \sigma^\ast} \,\,
(g_{\sigma i} \, \delta\sigma + g_{\sigma^\ast i} \, \delta\sigma^\ast)
\nonumber\\
+\sum_i g_{\sigma i} \,
(m_i/m_i^\ast-g_{\sigma i} \, \sigma/m_i^\ast
- g_{\sigma^* i} \, \sigma^*/m_i^\ast) \,
\delta n_i,
\label{sigma5}\\
m_{\sigma^\ast}^{2}  \, \delta \sigma^\ast
= -\sum_i g_{\sigma^\ast i} \, \frac{\partial R_i(x, 0)}{\partial x}
|_{x=m_i-g_{\sigma i} \, \sigma - g_{\sigma^\ast i} \, \sigma^\ast} \,\,
(g_{\sigma i} \, \delta\sigma + g_{\sigma^\ast i} \, \delta\sigma^\ast)
\nonumber\\
+ \sum_i {g_{\sigma^\ast i} \,
(m_i/m_i^\ast-g_{\sigma i} \, \sigma/m_i^\ast -
g_{\sigma^* i} \, \sigma^*/m_i^\ast)}\,
\delta n_i.
\label{sigma*5}
\end{eqnarray}
In the matrix form Eqs.\ (\ref{sigma5}) and (\ref{sigma*5})
can be rewritten as
\begin{eqnarray}
&&\left(
\begin{array}{ccc}
m_{\sigma}^2+2 \, g_3 \, \sigma + 3 \, g_4 \, \sigma^2 + I_{\sigma \sigma} &
 I_{\sigma \sigma^\ast}\\
I_{\sigma \sigma^\ast}&
m_{\sigma^\ast}^2 + I_{\sigma^\ast \sigma^\ast}
\end{array}
\right) \,
\left(
\begin{array}{ccc}
\delta \sigma \\
 \delta \sigma^\ast
\end{array}
\right) =
\nonumber\\
&=&
\left(
\begin{array}{ccc}
\sum_i {g_{\sigma i} \,
[m_i/m_i^\ast-g_{\sigma i} \, \sigma/m_i^\ast -
g_{\sigma^* i} \, \sigma^*/m_i^\ast]}\,
\delta n_i \\
 \sum_i {g_{\sigma^\ast i} \,
[m_i/m_i^\ast-g_{\sigma i} \, \sigma/m_i^\ast -
g_{\sigma^* i} \, \sigma^*/m_i^\ast]}\,
\delta n_i
\end{array}
\right),
\label{matrix2}
\end{eqnarray}
where we defined
\begin{eqnarray}
I_{\sigma \sigma} &=& \sum_i g_{\sigma i}^2 \, \frac{\partial R_i(x, 0)}{\partial x}
|_{x=m_i-g_{\sigma i} \, \sigma - g_{\sigma^\ast i} \, \sigma^\ast},
\label{Iss}\\
I_{\sigma \sigma^\ast} &=& \sum_i g_{\sigma i} \, g_{\sigma^\ast i} \, \frac{\partial R_i(x, 0)}{\partial x}
|_{x=m_i-g_{\sigma i} \, \sigma - g_{\sigma^\ast i} \, \sigma^\ast},
\label{Iss*}\\
I_{\sigma^\ast \sigma^\ast} &=& \sum_i g_{\sigma^\ast i}^2 \, \frac{\partial R_i(x, 0)}{\partial x}
|_{x=m_i-g_{\sigma i} \, \sigma - g_{\sigma^\ast i} \, \sigma^\ast}.
\label{Is*s*}
\end{eqnarray}
The solution to the system (\ref{matrix2}) can be easily found.
To calculate the quantities $\delta \omega^0$, $\delta \rho^0$, and $\delta \phi^0$
we have to linearize the corresponding Eqs.\ (\ref{omega})--(\ref{phi}).
The result is
\begin{eqnarray}
&&m_\omega^{\ast 2} \, \delta \omega^0
+ 2 \left(
A_{\omega} \, \omega^0 \delta \omega^0 + A_{\rho} \, \rho_3^0 \, \delta \rho_3^0
+ A_{\phi} \, \phi^0 \, \delta \phi^0
\right) \, \omega^0 = \sum_i g_{\omega i} \, \delta n_i,
\label{omega6} \\
&&m_\rho^{\ast 2} \, \delta \rho_3^0
+ 2 \left(
B_{\omega} \, \omega^0 \delta \omega^0 + B_{\rho} \, \rho_3^0 \, \delta \rho_3^0
+ B_{\phi} \, \phi^0 \, \delta \phi^0
\right) \, \rho_3^0 = \sum_i g_{\rho i} \, I_{3i} \, \delta n_i,
\label{rho6} \\
&&m_\phi^{\ast 2} \, \delta \phi^0
+ 2 \left(
C_{\omega} \, \omega^0 \delta \omega^0 + C_{\rho} \, \rho_3^0 \, \delta \rho_3^0
+ C_{\phi} \, \phi^0 \, \delta \phi^0
\right) \, \phi^0 = \sum_i g_{\phi i} \, \delta n_i,
\label{phi6}
\end{eqnarray}
where the meson effective masses 
$m_{\omega}^\ast$, $m_{\rho}^\ast$, and $m_{\phi}^\ast$ 
are given by Eqs.\ (\ref{momega})--(\ref{mphi}).
In the matrix form the system of equations (\ref{omega6})--(\ref{phi6})
is presented as
\begin{equation}
\left(
\begin{array}{ccc}
m_{\omega}^{\ast 2}+2 \, A_{\omega} \, (\omega^0)^2 &
 2 \, A_{\rho} \, \omega^0 \, \rho_3^0 &
 2 \, A_{\phi} \, \omega^0 \, \phi^0 \\
2 \, B_{\omega} \, \omega^0 \, \rho_3^0 &
m_{\rho}^{\ast 2}+2 \, B_{\rho} \, (\rho^0_3)^2 &
2 \, B_{\phi} \, \rho_3^0 \, \phi^0 \\
2 \, C_{\omega} \, \omega^0 \, \phi^0 &
2 \, C_{\rho} \, \rho_3^0 \, \phi^0 &
m_{\phi}^{\ast 2}+2 \, C_{\phi} \, (\phi^0)^2
\end{array}
\right)
\left(
\begin{array}{ccc}
\delta \omega^0 \\
 \delta \rho_3^0\\
\delta \phi^0
\end{array}
\right) =
\left(
\begin{array}{ccc}
\sum_i \, g_{\omega i} \, \delta n_i \\
 \sum_i \, g_{\rho i} \, I_{3i} \,\delta n_i\\
\sum_i \, g_{\phi i} \, \delta n_i
\end{array}
\right).
\label{matrix3}
\end{equation}
The solution to this matrix equation can also be easily obtained.
Schematically, expressions for $\delta \sigma$, $\delta \sigma^\ast$,
$\delta \omega^0$, $\delta \rho_3^0$, and $\delta \phi^0$ can be
written as
\begin{eqnarray}
\delta \sigma &=& \sum_j \beta_{\sigma j} \, \delta n_j,
\label{ds}\\
\delta \sigma^\ast &=& \sum_j \beta_{\sigma^\ast j} \, \delta n_j,
\label{ds*}\\
\delta \omega^0 &=& \sum_j \beta_{\omega j} \, \delta n_j,
\label{dw}\\
\delta \rho_3^0 &=& \sum_j \beta_{\rho j} \, \delta n_j,
\label{drho}\\
\delta \phi^0 &=& \sum_j \beta_{\phi j} \, \delta n_j,
\label{dphi}
\end{eqnarray}
where we assume that the quantities $\beta_{\sigma j},\ldots,\beta_{\phi j}$
have been already calculated from Eqs.\ (\ref{matrix2}) and (\ref{matrix3}).
Finally, 
taking into account Eqs.\ (\ref{ds})--(\ref{dphi}) and
comparing Eqs.\ (\ref{Ei}) and (\ref{EiLandau}), one finds
the following expression for the Landau parameters $f_0^{ij}$,
\begin{eqnarray}
f_0^{ij} &=& g_{\omega i} \, \beta_{\omega j}
+ g_{\rho i} \, I_{3i} \, \beta_{\rho j} + g_{\phi i} \, \beta_{\phi j}
\nonumber\\
&-& (m_i/m_i^\ast-g_{\sigma i} \, \sigma/m_i^\ast
- g_{\sigma^* i} \, \sigma^*/m_i^\ast)\,
(g_{\sigma i} \, \beta_{\sigma j} + g_{\sigma^\ast i} \, \beta_{\sigma^\ast j}).
\label{f0ij}
\end{eqnarray}

It can be shown (see, e.g.,  \citealt{gkh09a})
that these parameters
are directly related to
the derivatives $\partial \mu_i/\partial n_j$, 
which should be taken at fixed particle number densities $n_k$ ($k \neq j$).
Namely, one has the following relation
\begin{equation}
\frac{\partial \mu_i(n_{\rm n}, \ldots,\, n_{\Sigma^+})}{\partial n_j}
=\frac{\partial \mu_j(n_{\rm n}, \ldots,\, n_{\Sigma^+})}{\partial n_i}
= f_0^{ij}+\frac{1}{N_{{\rm F}i}} \delta_{ij},
\label{dmudnj}
\end{equation}
where $N_{{\rm F} i}$ is defined by Eq.\ (\ref{eq:F.f}).

\section{Numerical results}
\label{sect:results}

\subsection{RMF models, EOSs and \texorpdfstring{$M_{\rm max}$}{Lg}}
\label{sect:EOS.Mmax}

We consider  three RMF models of NH matter 
(we will call them {\bf GM1A}, {\bf GM$1^\prime$B} and {\bf TM1C}), 
which are
specific realizations of $\sigma\omega\rho\phi\sigma^\ast$ model
of \cite{bm09}.
The parameters of the models are given in the Appendix \ref{appendix:coupling}. 
Below we give their brief characteristics.
For all the models the binding energy
per nucleon at saturation is $B_{\rm s}=-16.3$~MeV.
Moreover, they all reproduce the semi-empirical depths of potential wells for
hyperons at rest in symmetric nuclear matter at saturation density,
$U^{({\rm N})}_\Lambda=-28~$MeV~, $U^{({\rm N})}_\Xi=-18~$MeV~, $U^{({\rm N})}_\Sigma=30~$MeV
(e.g., \citealt*{mdg88}). The parameters of NS configuration with
$M_{\rm max}$ for non-rotating NS models are given in Table \ref{tab:Mmax} 
for {\bf GM1A}, {\bf GM$1^\prime$B} and {\bf TM1C} models.

\begin{table}
\begin{center}
\begin{tabular}[t]{ccccccc}
\hline\hline
&&&&&&\\
 RMF model  &  $M_{\rm max}$   & $R(M_{\rm max})$ & $\rho_{\rm max}/10^{15}$  &
      $n_{\rm b, max}$ & $-\left(y_{\rm S}\right)_{\rm c}$   & $-\left<y_{\rm S}\right>$  \\
      &   $(\msun)$ &   $({\rm km})$ & $(\mdens)$ &  $(\bdens)$ &   &    \\
 \hline
 &&&&&&\\
  {\bf GM1A}  & 1.994  & 12.05  &  2.00   &  0.923  & 0.607  & 0.143    \\
    &&&&&&\\
  {\bf GM1$^\prime$B}  & 2.015  & 11.45 &  2.28  &   1.018 &  0.671  & 0.181 \\
    &&&&&&\\
  {\bf TM1C}   & 2.056  &  12.51 & 1.85  &   0.856  & 0.493   & 0.093 \\
    &&&&&&\\
 \hline\hline
\end{tabular}
\end{center}
\caption{Parameters of non-rotating NS models with maximum allowable mass. 
The columns are (from left to right): 
RMF model of NH matter, 
maximum stellar mass in units of the solar mass, 
corresponding radius of the star in km, 
central density in $\mdens$, 
central baryon number density in $\bdens$, 
the ratio of (minus) strangeness number density $S$ to baryon number density $n_{\rm b}$
[$y_{\rm S}=-S/n_{\rm b}=-(n_\Lambda+2n_{\Xi^-}+2n_{\Xi^0})/n_{\rm b}$] 
in the centre of the star, 
the same ratio but averaged over the whole star.
}
\label{tab:Mmax}
\end{table}

\vskip 2mm
\noindent
{\bf GM1A}. In the nucleon sector this is the {\bf GM1} model
of \cite{gm91}. 
The saturation baryon number density for this model $n_{\rm s}=0.153$~fm$^{-3}$.
Nuclear matter incompressibility at the saturation point,
$K_{\rm s}= 300$ MeV,
is somewhat  larger than the semi-empirical estimates of this quantity.
The symmetry energy $E_{\rm sym}=32.5$ MeV is  within the
semi-empirical evaluations.
The {\it Dirac effective nucleon mass} 
in symmetric nuclear matter at saturation is
$m_{{\rm D} \, {\rm s}}^\ast \equiv m_{\rm N}-g_{\sigma {\rm N}}\sigma=0.7 m_{{\rm N}}$, 
where  $m_{\rm N}\equiv (m_{\rm n}+m_{\rm p})/2 \approx 938.919$~MeV. 
The model is then
extended to the NH matter. The vector-meson coupling constants with hyperons are
obtained from the nucleon ones using the SU(6) symmetry. The scalar $\sigma^\ast$
meson is  not included. Inclusion of the vector $\phi$ meson producing repulsion
between hyperons  is sufficient to make the model (marginally) consistent with
$2.0~\msun$ pulsars.

\vskip 2mm
\noindent
{\bf GM$1^\prime$B}. 
The saturation baryon number density
is the same as for {\bf GM1A}, $n_{\rm s}=0.153$~fm$^{-3}$.
The nuclear  matter incompressibility 
at saturation point, 
$K_{\rm s}=240$ MeV, 
is within typical semi-empirical evaluations
of $K_{\rm s}$: it is significantly lower than that obtained for
the {\bf GM1A} model. In contrast, the symmetry energy  and the
Dirac effective nucleon mass  are  the same  as those obtained
for the {\bf GM1A} model.
In the NH matter, the vector $\phi$ meson  is included, while
the scalar $\sigma^\ast$ is not present.
In spite of a significantly lower $K_{\rm s}$, compared to that obtained for
{\bf GM1A},  the value of $M_{\rm max}$ is above
$2.0~\msun$. This is  due to a breaking  of the SU(6) symmetry in the
vector-meson couplings to hyperons. Using notation of \cite{wcs12b},
this symmetry breaking is  characterized by $z=0.3$, which
is significantly smaller than $z=1/\sqrt{6}\simeq 0.408$ corresponding to
the SU(6)-symmetric case.

\vskip 2mm
\noindent
{\bf TM1C}. 
It reduces to the widely used {\bf TM1} model in the nucleon  sector,
see \cite{st94}. 
For the latter model $n_{\rm s}=0.145$~fm$^{-3}$.
The  nuclear matter incompressibility,
$K_{\rm s}=281~$MeV is on the high-side of semi-empirical evaluations.
The Dirac effective nucleon mass in symmetric nuclear matter at saturation point is
rather small, $m_{{\rm D} \, {\rm s}}^\ast=0.634 m_{\rm N}$, 
where $m_{\rm N}$ is chosen to be $m_{\rm N} \equiv 938$~MeV for this model. 
The nuclear symmetry energy, $E_{\rm sym}=36.9~$MeV,
is higher than  typical semi-empirical evaluations.
Extension of {\bf TM1} to NH matter includes the vector $\phi$ meson
{\it and} the  scalar $\sigma^\ast$ meson.
The breaking of the SU(6) symmetry is even stronger than for the
{\bf GM$1^\prime$B} model, and corresponds to $z=0.2$.
In addition to fitting the $U_\Lambda^{({\rm N})}$, $U_\Xi^{({\rm N})}$,
and $U_\Sigma^{({\rm N})}$ potential well depths, this model also fits a
weak $\Lambda-\Lambda$ attraction, $U^{(\Lambda)}_\Lambda=-5.0~$MeV 
(\citealt{takahashi_et_al_01}),
and assumes $U^{(\Xi)}_\Xi \approx U^{(\Xi)}_\Lambda 
\approx 2 U_{\Xi}^{(\Lambda)} \approx
2U^{(\Lambda)}_\Lambda \approx -10.0~$MeV
(\citealt{sdggms94}). 
Maximum allowable
mass is $2.056~\msun$.

\vskip 2mm

The EOSs for these models,  $P=P(\rho)$, are plotted in Fig.\ \ref{fig:Prho}. 
The way they are obtained is briefly discussed 
in Appendix \ref{appendix:eos}. 
One notices that
for $\rho \ga 2\times 10^{15}~\mdens$ the EOS {\bf TM1C} is the softest
one. And still, it yields the highest value of $M_{\rm max}$. This apparent
paradox can be explained as follows. $M_{\rm max}$ is a functional of the EOS,
$M_{\rm max}[P(\rho<\rho_{\rm max})]$, but the EOS for $\rho$ greater than
the maximum central density in {\it stable} NSs does not affect
the value of $M_{\rm max}$. {\bf TM1C} is actually the stiffest for
$\rho \la 1.4\times 10^{15}~\mdens$, which is quite close to the maximum
central density $\rho_{\rm max}\approx1.85\times 10^{15}~\mdens$ for stable NS
based on this EOS.  
Therefore, while {\bf TM1C} is the softest
EOS for $\rho \ga 2\times 10^{15}~\mdens$, this is irrelevant for the value of
$M_{\rm max}$.

\begin{figure}
\setlength{\unitlength}{1mm} \leavevmode \hskip  0mm
\includegraphics[width=150mm,clip]{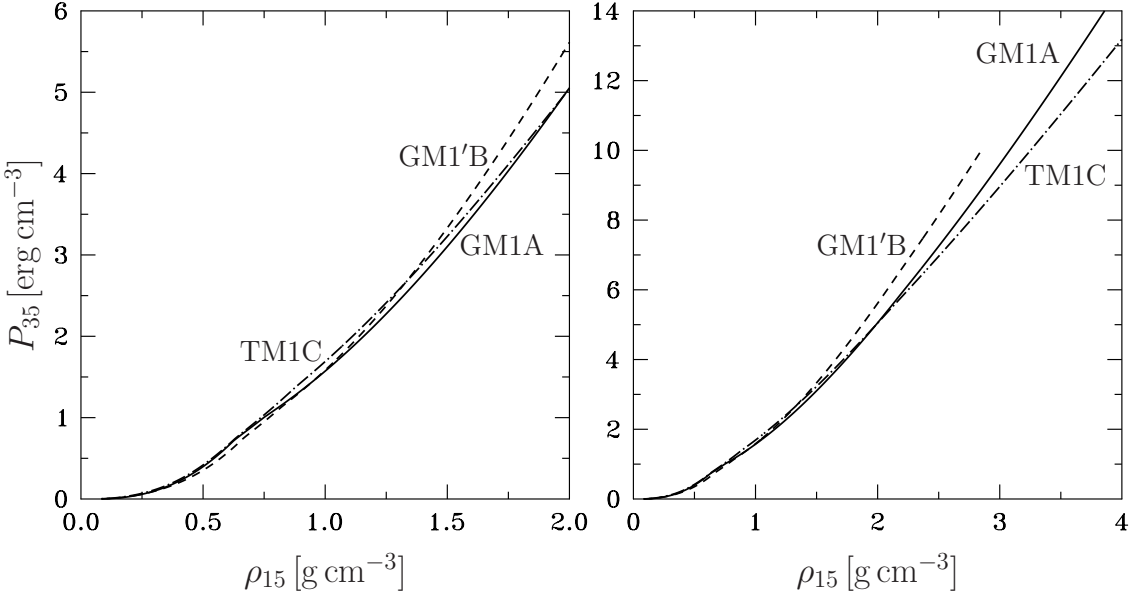}
\caption {Pressure $P_{35}=P/10^{35}$ versus density $\rho_{15}=\rho/10^{15}$ for
three models of NH matter considered in this paper. {\it Right-hand panel:}
Overall plots of EOSs. For $\rho_{15}>2$, the {\bf TM1C} EOS is the softest, and
{\bf GM$1^\prime$B} the stiffest. {\it Left-hand panel:} Lower-density, $\rho_{15}<2$, segments
of the EOSs. The ordering of the EOSs according to their stiffness depends on the
density interval. For further discussion of this effect and its impact on the value
of $M_{\rm max}$ see the text.}
\label{fig:Prho}
\end{figure}

\subsection{Particle fractions, adiabatic indices, and the speed of sound}
\label{sect:yi.vsound}

\begin{figure}
\setlength{\unitlength}{1mm} \leavevmode \hskip  0mm
\includegraphics[width=170mm,clip]{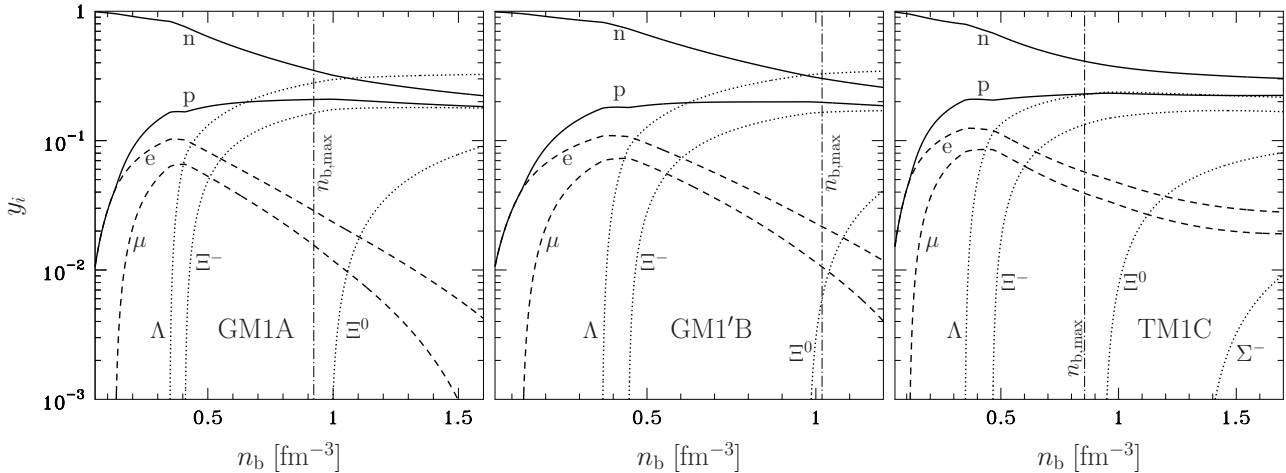}
\caption {
Particle fractions $y_i=n_i/n_{\rm b}$ versus baryon number density $n_{\rm b}$
for three EOSs, {\bf GM1A}, {\bf GM1$^\prime$B}, and {\bf TM1C}. 
The vertical dot-dashed lines correspond to the
maximum baryon number density reached in stable non-rotating
NSs for a given EOS, see Table \ref{tab:Mmax}. For further details
see Sect.\ \ref{sect:yi.vsound}.}
\label{fig:y_i-EOS}
\end{figure}

In Fig.\ \ref{fig:y_i-EOS} we show the particle fractions of constituents of NH matter,
$y_i\equiv n_i/n_{\rm b}$, as functions of baryon number density $n_{\rm b}$. 
Three panels correspond to three RMF models ({\bf GM1A}, {\bf GM$1^\prime$B}, and {\bf TM1C}).
Dot-dashed vertical lines correspond to the maximum baryon number density 
reachable in stable non-rotating NSs, see Table \ref{tab:Mmax}.
The order of
appearance of hyperons with increasing density is identical for all EOSs. 
The corresponding thresholds are presented in Table \ref{table:thresholds}. 
The first hyperon to appear is $\Lambda$,
the second hyperon is $\Xi^-$.
The third hyperon, $\Xi^0$, appears only in model {\bf GM$1^\prime$B} and 
exists only in configurations close to
the $M_{\rm max}$ one, thus playing  a marginal role in stable stars. 
A large repulsive potential energy of $\Sigma^-$ 
in nuclear matter makes its threshold density very high, from $9n_{0}$ for {\bf TM1C}
to more than $10n_{0}$ for {\bf GM1A}. 
Therefore, $\Sigma^-$ are absent in stable NSs.
 
\begin{table}
\begin{center}
\begin{tabular}[t]{cccccc}
\hline\hline
 Model  &  $n_{\rm b}^{(\mu)}$   & $n_{\rm b}^{(\Lambda)}$ & $n_{\rm b}^{(\Xi^-)}$  & $n_{\rm b}^{(\Xi^0)}$ & $n_{\rm b, max}$\\
      &   $({\rm fm}^{-3})$ &   (${\rm fm}^{-3}$) & (${\rm fm}^{-3}$) &  (${\rm fm}^{-3}$)& (${\rm fm}^{-3}$)     \\
 \hline
{\bf GM1A}  & 0.1271  & 0.3472  & 0.4076 & -- & 0.923  \\
    {\bf GM1$^\prime$B}  & 0.1272  & 0.3669 &  0.4438  &   0.9750 & 1.018 \\
{\bf TM1C}   & 0.1090  & 0.3466  &  0.4622 &  --  & 0.856 \\
\hline\hline
\end{tabular}
\end{center}
\caption{Thresholds $n_{\rm b}^{(k)}$ of appearance
of particles $k=\mu$, $\Lambda$, $\Xi^-$, and $\Xi^0$ 
for which $n_{\rm b}^{(k)}<n_{\rm b, max}$ (see the last column). 
Only model {\bf GM1$^\prime$B} admits the existence of 
$\Xi^0$ hyperons in stable NSs.}
\label{table:thresholds}
\end{table}

An important quantity characterizing dynamic 
properties of stellar matter is the adiabatic index
\begin{equation}
\gamma=\frac{P+\rho}{P}\,\,\frac{\delta P}{\delta \rho},
\label{gammagamma}
\end{equation}
where $\delta P$ is a small deviation of the pressure $P$ from its equilibrium value 
caused by a small variation $\delta \rho$ of the energy density $\rho$. 
This index is related to the speed of sound $s$
by the equality $s=\left[\gamma P/(P+\rho)\right]^{1/2}$. 
The ratio
$\delta P/\delta \rho$ in Eq.\ (\ref{gammagamma})
should be calculated 
under a number of additional conditions 
(such as quasineutrality, chemical equilibrium etc.),
which differ depending on a timescale $\tau$ of a physical process under consideration
\footnote{The most natural example of such process is the NS oscillations.
Then $\tau \sim 1/\omega$, where $\omega$ is the oscillation frequency.}.
The resulting adiabatic indices $\gamma$ will also be different.

\begin{figure}
\setlength{\unitlength}{1mm} \leavevmode \hskip  0mm
\includegraphics[width=170mm,clip]{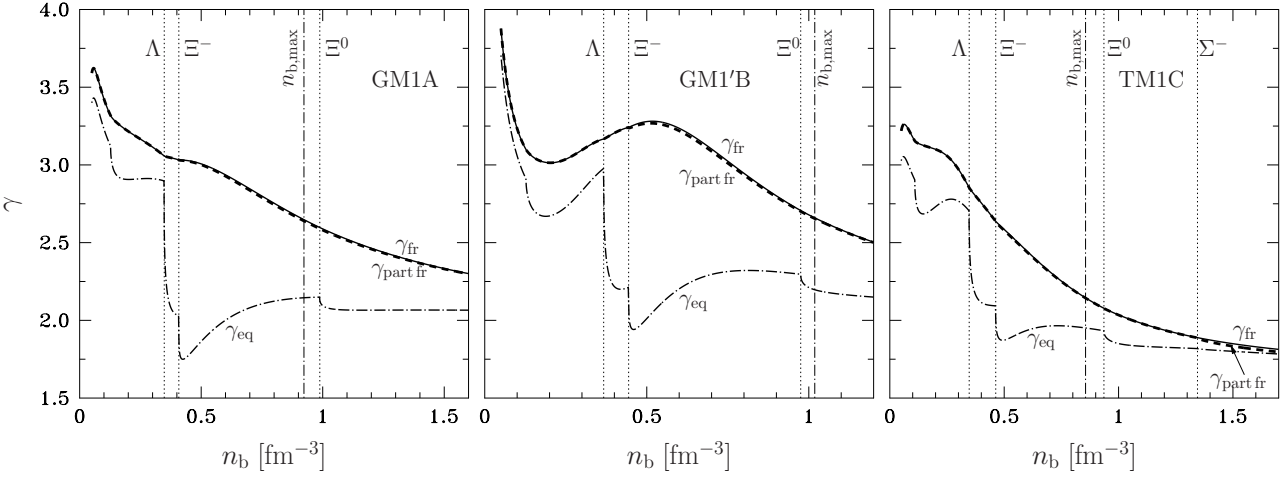}
\caption {
Three adiabatic indices versus $n_{\rm b}$ for the three selected models of NH matter.
$\gamma_{\rm eq}$ (dot-dashed lines) is calculated assuming full thermodynamic equilibrium;
$\gamma_{\rm fr}$ (solid lines) is obtained under assumption 
that all reactions of particle mutual transformations are frozen 
(completely frozen matter composition);
$\gamma_{\rm part\,fr}$ (dashed lines) 
assumes equilibrium with respect to the `fast' reactions, 
while `slow' reactions are frozen 
(see Sect.\ \ref{sect:yi.vsound} for details).
Vertical dotted lines indicate thresholds of appearance of (from left to right)
$\Lambda$, $\Xi^-$, $\Xi^0$, and $\Sigma^{-}$ hyperons. 
Vertical dot-dashed lines
show the maximum baryon number density 
in a non-rotating NS of a maximum allowable mass (see also Table \ref{tab:Mmax}). 
}
\label{fig:adiabat}
\end{figure}

Here we consider three adiabatic indices:
equilibrium adiabatic index $\gamma_{\rm eq}$ (\citealt*{hpy07}), 
frozen adiabatic index $\gamma_{\rm fr}$ (\citealt{hpy07}), 
and 
`partly frozen' adiabatic index $\gamma_{\rm part \,fr}$.
In Fig.\ \ref{fig:adiabat} they 
are shown by, respectively, 
dot-dashed, solid, and dashed lines as functions of $n_{\rm b}$ 
for the three models of NH matter adopted in this paper.

The index $\gamma_{\rm eq}$ naturally appears in the situation 
when the dynamical process of interest is very slow.
This means that
$\tau \gg \tau_{\rm strong}$ and $\tau \gg \tau_{\rm weak}$, 
where $\tau_{\rm strong}$ and $\tau_{\rm weak}$ 
are the characteristic timescales of
`fast' (due to strong interaction) and `slow' (due to weak interaction) 
reactions of particle mutual transformations,
which move the system towards full thermodynamic equilibrium 
(see, e.g., \citealt{ykgh01,kg09}).

The index $\gamma_{\rm fr}$ can be introduced (e.g., \citealt{hpy07}) 
in the opposite limit,
when $\tau \ll \tau_{\rm strong}$ and $\tau \ll \tau_{\rm weak}$.
In that case 
the process is so fast that {\it all} 
the reactions 
are effectively `frozen' on a dynamical timescale $\tau$.
Mathematically, this means that the particle fractions $y_i$
remain constant
during this process for {\it any} particle species 
$i={\rm e},\, \mu,\, {\rm n}, \,  {\rm p} ,\, \Lambda,\, \ldots $:
$y_i=n_i/n_{\rm b}={\rm constant}$.

Finally, the index $\gamma_{\rm part \, fr}$
is introduced in the intermediate case,
when $\tau_{\rm weak} \gg \tau \gg \tau_{\rm strong}$.
In that case the matter is in equilibrium with respect 
to the fast reactions, while slow reactions 
(such as, e.g., Urca reactions; see \citealt{hpy07}) 
are frozen.
In stable NSs the fast reactions are:
${\rm p}+\Xi^- \leftrightarrow \Lambda + \Lambda$ 
and ${\rm n}+\Xi^0 \leftrightarrow \Lambda + \Lambda$. 
The adiabatic indices 
are considered in more detail in Appendix \ref{app:adiabat}.

As follows from Fig.\ \ref{fig:adiabat},  
each time, when a hyperon species appears
as $n_{\rm b}$ increases, 
we see 
a sharp drop of the equilibrium adiabatic index $\gamma_{\rm eq}$.
Such drops reflect the fact that appearance of hyperons makes the EOS softer.
The magnitude of the hyperon-threshold drops decreases with increasing density, 
with the largest drop at the threshold for $\Lambda$'s. 
This is not surprising and is related to the increasing number of baryon species with growing density
(the more the baryon species, the less sensitive is $\gamma_{\rm eq}$ 
to the appearance of additional hyperon species). 

In contrast to $\gamma_{\rm eq}$, $\gamma_{\rm fr}$ 
(and $\gamma_{\rm part \, fr}$, 
which is practically indistinguishable from $\gamma_{\rm fr}$) 
does not drop sharply near the hyperon thresholds:
The influence of hyperon thresholds is less pronounced 
if we consider rapid processes with 
$\tau \ll \tau_{\rm strong}$ and $\tau \ll \tau_{\rm weak}$.

\begin{figure}
\setlength{\unitlength}{1mm} \leavevmode \hskip  0mm
\includegraphics[width=95mm,clip]{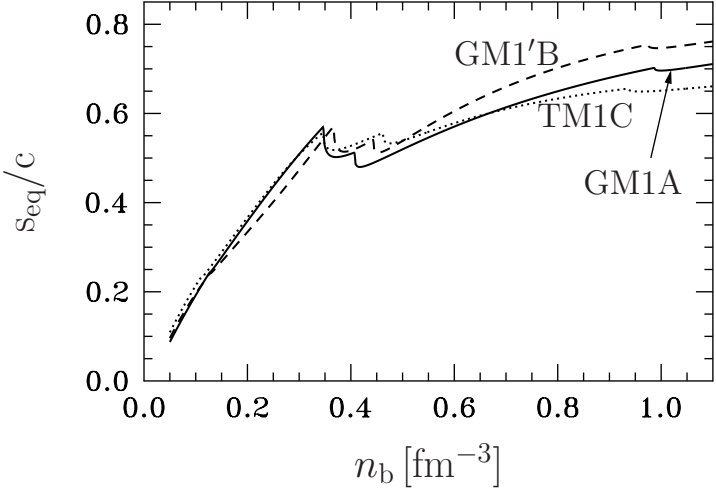}
\caption {
Equilibrium sound speed $s_{\rm eq}=[\gamma_{\rm eq} P/(P+\rho)]^{1/2}$ 
(in units of speed of light $c$) 
versus $n_{\rm b}$ for three models of NH matter.
}
\label{fig:sound}
\end{figure}

For illustration, Fig.\ \ref{fig:sound} shows the equilibrium speed of sound 
$s_{\rm eq}=\left[\gamma_{\rm eq} P/(P+\rho)\right]^{1/2}$ 
for the three EOSs considered in this paper. 

\subsection{Effective masses}
\label{sect:m.eff}

\begin{figure}
\setlength{\unitlength}{1mm} \leavevmode \hskip  0mm
\includegraphics[width=170mm,clip]{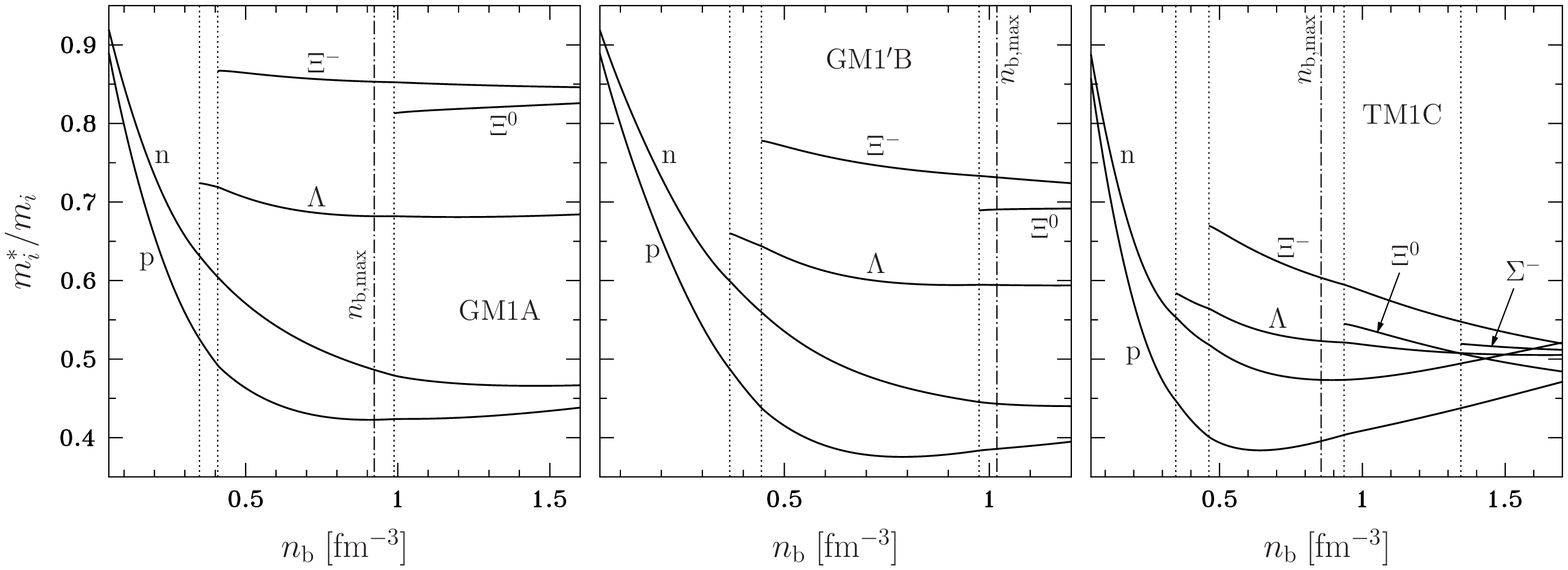}
\caption {The normalized Landau effective masses $m^{\ast}_i/m_i$ versus $n_{\rm b}$ for three  EOSs.
Each curve is marked by a corresponding baryon species index $i={\rm n}$, ${\rm p}$, $\Lambda$, $\ldots$. 
Other notations are the same as in Fig.\ \ref{fig:adiabat}.}
\label{fig:m_eff-EOS}
\end{figure}

Our results for normalized Landau effective masses ${\overline m}^\ast_i\equiv
m^\ast_i/m_i$ [see Eq.\ (\ref{meffLandau})] are shown in Fig.\ \ref{fig:m_eff-EOS}.

For all EOSs at all densities  ${\overline m}^\ast_{\rm n}>{\overline m}^\ast_{\rm p}$, 
and, moreover, at densities relevant to stable NSs 
$\overline{m}_{\Xi^-}^{\ast}>\overline{m}_{\Xi^0}^{\ast}>\overline{m}_\Lambda^{\ast}
>\overline{m}_{\rm n}^{\ast}>\overline{m}_{\rm p}^{\ast}$. 
 At the same time one notices a systematic differences
in ${\overline m}^\ast_{i_{\rm _H}}(n_{\rm b})$ curves 
(hereafter $i_{\rm _H}=\Lambda, \Xi^-, \Xi^0, \Sigma^-$) between three
dense matter models used. 

For {\bf GM1A} all hyperon ${\overline m}^\ast_{i_{\rm _H}}$ curves are
very flat, ranging from 
$0.7$ to $0.85$.
For {\bf GM$1^\prime$B} model, the values of
${\overline m}^\ast_{i_{\rm _H}}$ are systematically smaller,  about 
$0.6-0.75$.
The strongest Fermi-liquid effect and density
dependence are obtained for {\bf TM1C} model,
${\overline m}^\ast_{i_{\rm _H}}$ fall into the range $0.5-0.65$.

We conclude that there is a significant model
dependence of ${\overline m}^\ast_i(n_{\rm b})$ for hyperons. 
This fact reflects
limitations of our knowledge of the N--H and H--H interactions in dense NH matter.

\begin{figure}
\setlength{\unitlength}{1mm} \leavevmode \hskip  0mm
\includegraphics[width=170mm,clip]{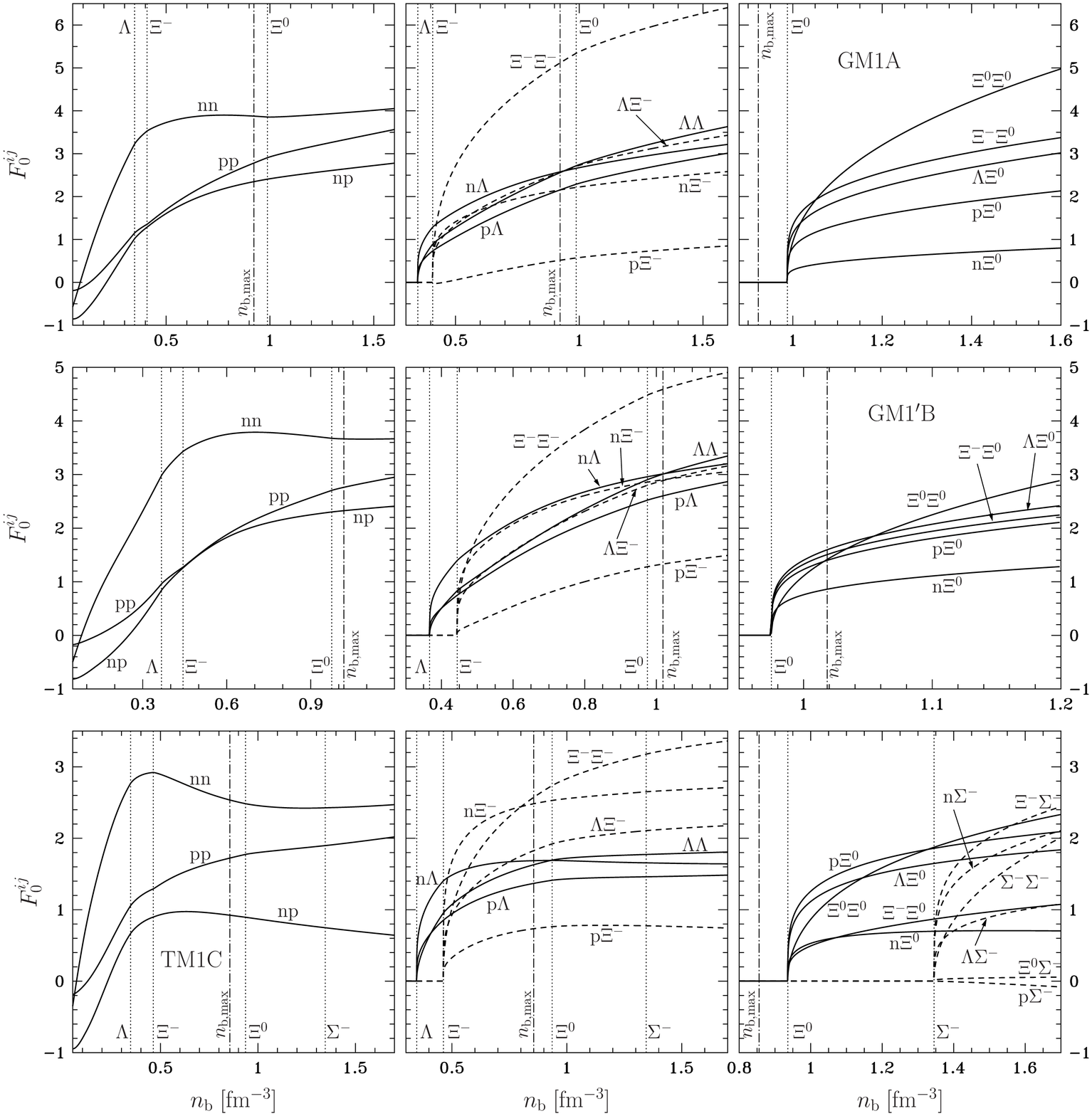}
\caption { The dimensionless Landau Fermi-liquid parameters $F^{ij}_0$ versus baryon number 
density for  {\bf GM1A} (upper panels), {\bf GM$1^\prime$B} (middle panels), and {\bf TM1C} (bottom
panels) RMF models. Each curve is marked by the
corresponding symbol $ij$. 
Other notations are the same as in Fig.\ \ref{fig:adiabat}.}
\label{fig:F0ik}
\end{figure}

\begin{figure}
\setlength{\unitlength}{1mm} \leavevmode \hskip  0mm
\includegraphics[width=170mm,clip]{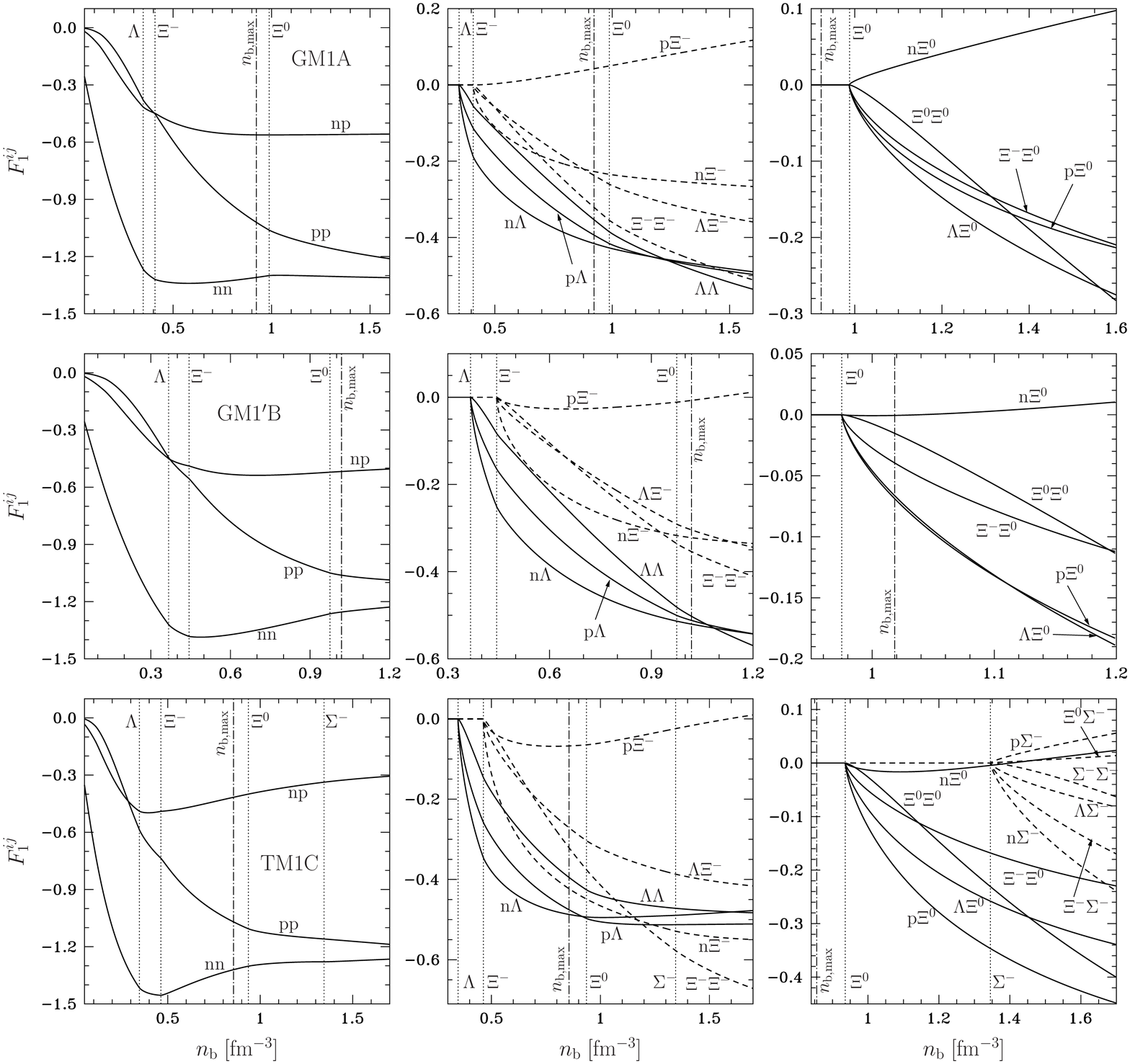}
\caption { 
The same as Fig.\ \ref{fig:F0ik}, but for $F^{ij}_1$.
}
\label{fig:F1ik}
\end{figure}

\begin{figure}
\setlength{\unitlength}{1mm} \leavevmode \hskip  0mm
\includegraphics[width=150mm,clip]{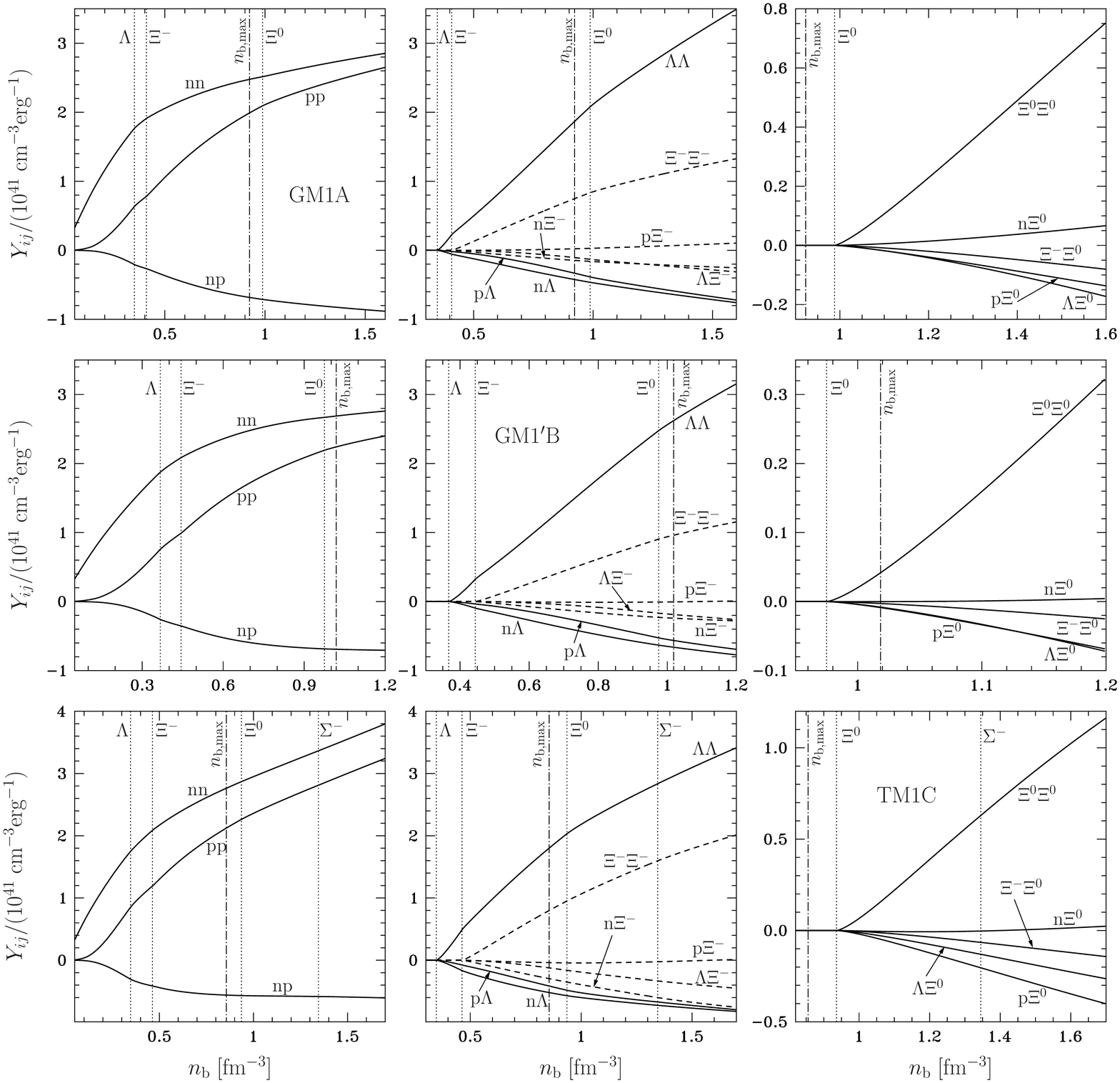}
\caption { 
The same as Fig.\ \ref{fig:F0ik}, but for 
normalized elements of symmetric matrix $Y_{ij}$.
For readability of the figure, 
we do not plot bunch of curves corresponding to the elements $Y_{i \Sigma^-}$
of the entrainment matrix.
}
\label{fig:Yik}
\end{figure}

\subsection{Landau Fermi-liquid parameters \texorpdfstring{$f_0^{ij}$}{Lg}, \texorpdfstring{$f_1^{ij}$}{Lg},
and entrainment matrix \texorpdfstring{$Y_{ij}$}{Lg}}
\label{sect:F0}

Our results for dimensionless Landau Fermi-liquid parameters $F^{ij}_0=F^{ji}_0$ and $F^{ij}_1=F^{ji}_1$ 
are collected in Figs.\ \ref{fig:F0ik} and \ref{fig:F1ik}, respectively.
The results for normalized dimensionless entrainment matrix $Y_{ij}=Y_{ji}$ are shown in
Fig.\ \ref{fig:Yik}.

Landau parameters $F_0^{i_{\rm N}j_{\rm N}}$ 
(hereafter $i_{\rm N}= {\rm n}, \, {\rm p}$) 
for {\bf GM}-type models have similar values and density dependence, those
for {\bf TM1C} model are smaller and are more similar to \cite{gkh09a} results. 
In contrast to \cite{gkh09a},
Landau parameters $F_0^{ij_{\rm H}}$ are all positive except for $F_0^{{\rm p}\Sigma^-}$ in the model {\bf TM1C} 
and $F_0^{{\rm p}\Xi^-}$ in the model {\bf GM1A} near the threshold for the $\Xi^-$ hyperon. 

Nucleon $\ell=1$ Landau parameters $F_1^{i_{\rm N}j_{\rm N}}$ are not so much model sensitive,
they are quite similar for the three models developed here and also for the model of \cite{gkh09a}.
We find that, as a rule, the $\ell=1$ Landau parameters are negative. 
A few ones 
which are positive, remain very small. 
$F_1^{\rm nn}$ dominates in magnitude over remaining Landau parameters, but $F^{\rm pp}_1$
becomes comparable to it at the largest densities. 
Model dependence of Landau parameters with
$i_{\rm N}j_{\rm H}$ and $i_{\rm H}j_{\rm H}$ indices
is more significant.

The entrainment matrix elements  $Y_{ij}$ are also not very model dependent.
The bundle of $Y_{ij_{\rm H}}$ is
bound by $Y_{\Lambda\Lambda}$ from above and by (negative) $Y_{{\rm n}\Lambda}$ from
below. $Y_{\Xi^-\Xi^-}$ is significantly smaller than $Y_{\Lambda\Lambda}$.
Non-diagonal matrix elements $Y_{ij}$ with $i\neq j$ 
are significantly smaller than $Y_{ii}$
or $Y_{jj}$, and are usually negative; if positive,  they are close to zero. 
On the opposite, diagonal elements $Y_{ii}$ are positive.

\subsection{Stability with  respect to \texorpdfstring{$\ell=0$}{Lg}  and \texorpdfstring{$\ell=1$}{Lg}
deformations of the Fermi surfaces}
\label{sect:stab.F0.F1}

A small $\ell=0$ deformation of the $i$-Fermi surface induces a small
perturbation $\delta n_i$ of particle number density $n_i$, and vice versa. 
We consider only long-wave (uniform) perturbations
that preserve 
electric charge neutrality of the system 
(in order to exclude the stabilizing effect of the Coulomb energy;
see \citealt{gkh09a} for details). 
Then the stability requirement is
equivalent to the positive definiteness of the quadratic form
$\sum_{k m}A_{km}\delta n_k\delta n_m$, 
where the indices $k$ and $m$ run over all particle species, 
except for electrons. 
The matrix $A_{km}$ is expressible in terms of the Fermi-liquid parameters $F^{ij}_0$
(see, e.g., \citealt{gkh09a}, and references therein). 
Stability with respect to perturbation $\delta n_i$ imposes a number of conditions on $F^{ij}_0$. 
We checked that these stability conditions 
are satisfied within the liquid NS core, i.e., for $n_{\rm b}>0.1~\bdens$, 
for all considered models.

A small $\ell=1$ deformation of the $i$-superfluid Fermi surface keeps the value of
$n_i$ unchanged but induces a uniform superfluid current associated with ${\pmb
Q}_i$. The change of the energy density associated with superfluid currents is
given by a quadratic form $\frac{1}{2}\sum_{ij}Y_{ij}{\pmb Q}_i{\pmb Q}_j$, see
\cite{gkh09a}. Stability of the ground state is equivalent to the positive
definiteness of the matrix $Y_{ij}$, 
implying a number of conditions on the parameters $F^{ij}_1$. 
We checked that these conditions 
are satisfied for all the three models 
and at all densities relevant to NSs.

Recently, \cite*{gro12} and \cite{grom13} pointed out the
first-order phase transition associated with the appearance of strangeness in dense
baryonic matter. This first-order phase transition is signalled by a spinodal
instability of an  uniform  baryon matter (${\rm n}\Lambda$-matter in \citealt{gro12},
${\rm np}\Lambda {\rm e}$-matter in \citealt{grom13}). In our calculations, the matrix $A_{ij}$ of
Sect.\ \ref{sect:stab.F0.F1} is positive definite at $n_{\rm b}>0.1~\bdens$ and appearance of $\Lambda$
is continuous (second-order phase transition): we do not find spinodal instability
associated with appearance of strangeness that would indicate a phase-separation
instability. However, in contrast to \cite{gro12,grom13} we consider
exclusively baryon matter with no trapped neutrinos and close to beta equilibrium.
Therefore, our  particle fractions $y_i$ 
{\it in equilibrium} result from the weak
interaction equilibrium conditions (see Eq.\ \ref{mucond}), 
and are functions of $n_{\rm b}$, $y_j=y_j^{\rm (eq)}(n_{\rm b})$. 
Our trajectory $y_j=y_j^{\rm (eq)}(n_{\rm b})$ does not cross a
spinodal instability region.

\section{Summary of results}
\label{sect:discussion}

We develop a general scheme for calculation of the $\ell=0,1$ Landau Fermi-liquid
parameters, valid for a broad class of nonlinear RMF
models of dense baryon matter.
A nonlinear Lagrangian that we consider involves the octet of baryons coupled to
the $\sigma\omega\rho\phi\sigma^\ast$ mesons. 
It includes quartic terms in meson fields.

Knowledge of the Landau Fermi-liquid parameters is crucial
for modelling NSs
because it allows one to directly calculate the following important quantities:
($i$) the thermodynamic derivatives $\partial \mu_i/\partial n_j$ 
[see Eqs.\ (\ref{dmudnj}) and (\ref{dmudnj2})], 
where $\mu_i$ and $n_j$ are the relativistic chemical potential
and the number density of particle species $i$ and $j$, respectively;
($ii$) the relativistic entrainment matrix $Y_{ij}$, 
both at zero temperature 
[see Eq.\ (\ref{eq:Y.F})]
and at finite temperatures (see \citealt{gkh09b});
this is a basic parameter for superfluid NSs.

The developed general scheme has been applied to 
study in detail three up-to-date specific RMF models of NH matter,
which are consistent
with the existence of $2~\msun$ pulsars 
(PSR J1614-2230 and J0348+0432; see 
\citealt{demorest_et_al_10, antoniadis_et_al_13})
and with semi-empirical nuclear and hypernuclear data.

These models allow for the presence of (maximum) three hyperon species 
in stable NSs. 
Two of the models ({\bf GM1A} and {\bf GM1$^\prime$B})
predict the appearance of (with increasing density)
$\Lambda$ and $\Xi^-$ hyperons, 
while the model {\bf TM1C} predicts also appearance of $\Xi^0$ hyperons
close to a maximum density reachable in stable NSs
for this model.
It is interesting that, in contrast to, e.g., the paper by \cite{gkh09a},
$\Sigma^-$ hyperons do not appear in stable NSs for the selected RMF models
because of their large repulsive potential energy in nuclear matter.

For all models we calculated 
and analysed 
the Landau Fermi-liquid parameters 
$F_{0}^{ij}$ and $F_1^{ij}$ 
as functions of the baryon number density $n_{\rm b}$, 
entrainment matrix $Y_{ij}(n_{\rm b})$ at $T=0$,
EOS [pressure versus density relation $P(\rho)$], 
particle number densities $n_i(n_{\rm b})$,
adiabatic indices, and Landau effective masses.

All obtained numerical results for the three RMF models constructed by us 
are available on-line as a public domain at: 
{\bf http://www.ioffe.ru/astro/NSG/heos/hyp.html}.
This data source contains all necessary information to model dynamics of superfluid NSs, 
e.g., their oscillations and cooling.
The description of the on-line material is presented in the Appendix \ref{appendix:online}.

\section*{Acknowledgements}

This work was partially supported by RF president programme 
(grants MK-857.2012.2 and NSh-4035.2012.2), by RFBR (grants
11-02-00253-a and 12-02-31270-mol-a), 
by the Ministry of Education and Science of
Russian Federation (agreement no.\ 8409, 2012),
and by the Polish 
NCN research grant no.\ 2011/01/B/ST9/04838.

\bibliography{literature}

\appendix

\section{Coupling constants}
\label{appendix:coupling}

Here we discuss the coupling constants for models 
{\bf GM1A}, {\bf GM1$^\prime$B}, and {\bf TM1C}.
The main parameters characterizing
these models are summarized in Table\ \ref{table:param}.
The actual values of hyperon and meson masses which were used 
in all calculations are presented in Table\ \ref{table:masses}
(note that the masses of baryons in each isomultiplet 
are assumed to be the same). 
For all the models $\Lambda_{\rm V}=\widetilde{c}_3=0$.
The data which are not included in Table\ \ref{table:param}
are the depths of potential wells for hyperons in symmetric nuclear matter at 
saturation density (e.g., \citealt{mdg88}; \citealt{sg00}; \citealt*{syt06}; \citealt{wcs12a}),
\begin{equation} 
U^{({\rm N})}_\Lambda=-28.0\,\, {\rm MeV}, 
\quad 
U^{({\rm N})}_\Xi=-18.0 \,\,{\rm MeV}, 
\quad
U^{({\rm N})}_\Sigma=30.0 \,\,{\rm MeV},
\label{UN}
\end{equation}
which are the same for all three models.
In addition, the model {\bf TM1C}, 
which allows for the presence of $\sigma^\ast$ meson, 
fits also the weak $\Lambda-\Lambda$ attraction 
(\citealt{takahashi_et_al_01}),
\begin{equation}
U^{(\Lambda)}_\Lambda=-5.0 \,\, {\rm MeV},
\label{ULL}
\end{equation}
and assumes (\citealt{sdggms94})
\begin{equation}
U^{(\Xi)}_\Xi \approx U^{(\Xi)}_\Lambda \approx 2 U_{\Xi}^{(\Lambda)} 
\approx 2U^{(\Lambda)}_\Lambda=-10.0 \,\, {\rm MeV}.
\label{UXIXI}
\end{equation}

Using these data, one can calculate various coupling constants for the models
{\bf GM1A}, {\bf GM1$^\prime$B}, and {\bf TM1C}.
Most of them are listed in Table\ \ref{table:const}.
The remaining constants are related to these from Table\ \ref{table:const}
by the following conditions (see, e.g., equation (11) of  \citealt{wcs12b})
\begin{eqnarray}
g_{\omega \Lambda} &=& \frac{\sqrt{2}}{\sqrt{2}+\sqrt{3} z} \, g_{\omega {\rm N}},
\,\,\,
g_{\omega \Sigma} = g_{\omega \Lambda},
\,\,\,
g_{\omega \Xi} = \frac{\sqrt{2}-\sqrt{3} \, z}{\sqrt{2}+\sqrt{3} \, z} \, g_{\omega {\rm N}},
\label{gwn}\\
g_{\rho \Lambda} &=& 0,
\,\,\,
g_{\rho \Sigma} = g_{\rho {\rm N}},
\,\,\,
g_{\rho \Xi} = g_{\rho {\rm N}},
\label{grn}\\
g_{\phi {\rm N}} &=& \frac{\sqrt{6} \, z - 1}{\sqrt{2}+\sqrt{3} \, z} \, g_{\omega {\rm N}},
\,\,\,
g_{\phi \Lambda}=g_{\phi \Sigma}= - \frac{1}{\sqrt{2}+\sqrt{3} \, z}\, g_{\omega {\rm N}},
\,\,\,
g_{\phi \Xi}= - \frac{1 + \sqrt{6} \, z}{\sqrt{2} + \sqrt{3} \, z} \, g_{\omega {\rm N}}.
\label{gfn}
\end{eqnarray}

Let us briefly describe how we calculated the coupling constants 
presented in Table\ \ref{table:const}.
The constants $g_3$, $g_4$, $g_{\sigma {\rm N}}$, $g_{\omega {\rm N}}$, and $g_{\rho {\rm N}}$
can be 
expressed through 
the parameters 
$n_{\rm s}$, $B_{\rm s}$, $E_{\rm sym}$, $K_{\rm s}$, and $m_{\rm D \, s}^\ast$.
The corresponding consideration is similar to that presented 
in section 4.8 of \cite{glendenning00}
with the only exception that in our case $\phi$ 
can be non-vanishing even for pure nucleon matter 
[because $g_{\phi {\rm N}}$ is nonzero and is related to $g_{\omega {\rm N}}$ by Eq.\ (\ref{gfn})].

The constants $g_{\sigma i}$, where $i=\Lambda$, $\Sigma$, and $\Xi$, 
can be obtained from the requirement that the energy of an $i$-hyperon 
(with the momentum ${\pmb p}=0$) 
in the symmetric nuclear matter {\it at saturation} is equal to $U_{i}^{({\rm N})}$, that is
\begin{equation}
m_i+U_{i}^{({\rm N})}=g_{\omega i} \omega^0+g_{\phi i} \phi^0+(m_i- g_{\sigma i} \sigma).
\label{gsn}
\end{equation}
To use this formula one should first calculate the fields 
$\sigma$, $\omega^0$, and $\phi^0$ using, respectively, 
Eqs.\ (\ref{sigma}), (\ref{omega}), and (\ref{phi}). 
Note that nucleons do not generate the $\sigma^\ast$-field,
and $\rho_3^0=0$ in symmetric nuclear matter.

Finally, following  \cite{ys08}, 
the constants $g_{\sigma^\ast i}$ ($i=\Lambda$, $\Xi$) 
for the model {\bf TM1C} 
are calculated assuming $U_{i}^{(\Xi)}=-10$~MeV 
[see Eq.\ (\ref{UXIXI})]. 
To calculate them
we consider symmetric matter at $n=n_{\rm s}$ composed of 
equal number of $\Xi^{-}$ and $\Xi^0$ hyperons, 
and add one more $i$-hyperon with momentum ${\pmb p}=0$ to this system. 
Then the energy of this $i$-hyperon will be
\begin{equation}
m_i + U_{i}^{(\Xi)}=g_{\omega i} \omega^0+g_{\phi i} \phi^0
+(m_i- g_{\sigma i} \sigma-g_{\sigma^\ast i } \sigma^\ast),
\label{gsast}
\end{equation}
so that $g_{\sigma^\ast i}$ can be easily found
from this formula provided that the fields 
$\sigma$, $\sigma^\ast$, $\omega^0$, $\phi^0$ are already calculated 
from Eqs.\ (\ref{sigma})--(\ref{omega}) and (\ref{phi}). 
Note that $\rho_3^0=0$ in symmetric matter is composed of $\Xi^{-}$ and $\Xi^0$ hyperons.
The constant $g_{\sigma^\ast \Sigma}$ is set equal to $g_{\sigma^\ast \Lambda}$.

\begin{table}
\begin{center}
\begin{tabular}[t]{cccccccccc}
\hline\hline
Model  &  $m_{\rm N}$ & $n_{\rm s}$ &  $B_{\rm s}$ & $E_{\rm sym}$ & $K_{\rm s}$ &  $m_{{\rm D} \, {\rm s}}^\ast/m_{\rm N}$  & $c_3$ & $z$ &  $\sigma^\ast$ \\   
     &  $({\rm MeV})$ &  $({\rm fm}^{-3})$ & $({\rm MeV})$ &  $({\rm MeV})$ & $({\rm MeV})$ & & & &    \\
\hline
{\bf GM1A} \,  & 938.919  & 0.153  &  -16.3   &  32.5  & 300.0  & 0.7  & 0.0 \,\, & $1/\sqrt{6}$ \,\, &  No \\
\hline    
{\bf GM1$^\prime$B} \, & 938.919  & 0.153  &  -16.3   &  32.5  & 240.0  & 0.7  & 0.0 \,\, & 0.3 \,\, &  No \\
\hline
{\bf TM1C}   & 938.0  &  0.145 & -16.3  &   36.9  & 281.0   & 0.634 & 71.3075 \,\, &  0.2 \,\, & Yes \\
 \hline\hline
\end{tabular}
\end{center}
\caption{Various physical parameters for three models: {\bf GM1A}, {\bf GM1$^\prime$B}, and {\bf TM1C}.
In the table 
$m_{\rm N}$ is the nucleon mass; 
$n_{\rm s}$ the saturation density; 
$B_{\rm s}$ binding energy per nucleon;
$E_{\rm sym}$ the symmetry energy;
$K_{\rm s}$ nuclear matter incompressibility and
$m_{{\rm D} \, {\rm s}}^\ast/m_{\rm N}$ the Dirac effective mass in units of $m_{\rm N}$.
All these quantities are given at saturation point.
Further, $c_3$ is the coupling constant characterizing non-linear interaction of $\omega$-mesons;
$z$ is the parameter introduced in 
\citealt{wcs12b} to describe
deviation of a given model from the SU(6)-symmetric case (the SU(6) value is $z=1/\sqrt{6}$). 
Finally, the last column indicates that $\sigma^\ast$-mesons 
are only allowed for {\bf TM1C} model.
}
\label{table:param}
\end{table}

\begin{table}
\begin{center}
\begin{tabular}[t]{ccccccccc}
\hline\hline
Mass, &  $m_\Lambda$ & $m_\Sigma$ &  $m_\Xi$ & $m_\sigma$ & $m_\omega$ &  $m_\rho$  & $m_\phi$ & $m_{\sigma^\ast}$   \\   
\hline
MeV        & 1115.63 \, & 1193.12 \, &  1318.1 \,   &  511.198 \, & 783.0 \, & 770.0 \, & 1020.0 \, &  975.0 \, \\
\hline\hline
\end{tabular}
\end{center}
\caption{Masses (in MeV) of hyperons and mesons adopted in all calculations.}
\label{table:masses}
\end{table}

\begin{table}
\begin{center}
\begin{tabular}[t]{ccccccccccccc}
\hline\hline
Model  &  $g_3$ (fm$^{-1}$) & $g_4$ &  $g_{\sigma {\rm N}}$ & $g_{\omega {\rm N}}$ & $g_{\rho {\rm N}}$ &  $g_{\sigma \Lambda}$  & $g_{\sigma \Sigma}$ & $g_{\sigma \Xi}$ &  $g_{\sigma^{\ast} {\rm N}}$ &  $g_{\sigma^{\ast} \Lambda}$ & 
$g_{\sigma^{\ast} \Sigma}$ & $g_{\sigma^{\ast} \Xi}$ \\   
\hline
{\bf GM1A} \,  & 9.840  & -6.693 \, &  8.897 \,  &  10.617 \, & 8.198 \, & 5.435  & 3.603 & 2.844 &  -- & -- &-- & --\\
\hline
{\bf GM1$^\prime$B} \, & 16.324  & -31.985  &  9.096   &  10.558  & 8.198  & 6.241  &  4.368 &  4.275 & -- & -- & -- & --\\
\hline
{\bf TM1C}   & 7.684  &  -2.224 & 10.038  &  12.300  & 9.275 & 7.733 & 6.037 & 6.320 & 0.0 & 1.585 & 1.585 & 6.999 \\
 \hline\hline
\end{tabular}
\end{center}
\caption{Coupling constants for the models {\bf GM1A}, {\bf GM1$^\prime$B}, and {\bf TM1C}.}
\label{table:const}
\end{table}

\section{Equation of state}
\label{appendix:eos}

Here we consider beta-equilibrated NH matter.
The condition of beta-equilibrium implies the following 
relations between the relativistic chemical potentials (e.g., \citealt{hpy07}),
\begin{equation}
\mu_i=\mu_{\rm n}-q_i \mu_{\rm e}, \quad \mu_{\rm e}=\mu_\mu,
\label{mucond}
\end{equation}
where $q_i$ is the charge of baryon species $i$ in units of the proton charge.
These equations should be supplemented by the quasineutrality condition,
\begin{equation}
\sum_i q_i n_i -n_{\rm e}-n_\mu=0.
\label{quasi}
\end{equation}
Together with Eqs.\ (\ref{mui})--(\ref{pres})
and the field equations (\ref{sigma})--(\ref{phi}), 
these relations allow us to find all thermodynamic quantities 
as functions of baryon number density $n_{\rm b}$, 
as well as to determine the function $P(\rho)$.

\section{Adiabatic indices}
\label{app:adiabat}

Here we describe in more detail the calculation of the adiabatic indices 
$\gamma_{\rm eq}$, $\gamma_{\rm fr}$,  and $\gamma_{\rm part\, fr}$.
All the indices are given by Eq.\ (\ref{gammagamma}), 
which can be represented as
\begin{equation}
\gamma=\frac{n_{\rm b}}{P}\,\,\frac{\delta P}{\delta n_{\rm b}},
\label{gammagamma2}
\end{equation}
where we make use of the fact that in thermodynamic equilibrium 
$P+\rho=\mu_{\rm n} n_{\rm b}$ and that for small deviations from thermodynamic equilibrium 
$\delta \rho = \mu_{\rm n} \delta n_{\rm b}$ (see, e.g., \citealt{gusakov07,gk08}).

\noindent
($i$) {\it Equilibrium adiabatic index $\gamma_{\rm eq}$}.

\noindent
As it is discussed in Sec.\ \ref{sect:yi.vsound} 
in that case the ratio $\delta P/\delta n_{\rm b}$
should be calculated in full thermodynamic equilibrium, 
that is, under conditions (\ref{mucond}) and (\ref{quasi}).
In this situation, $P$ can be presented as only a function of $n_{\rm b}$,
while other particle number densities can be expressed through $n_{\rm b}$ 
by means of Eqs.\ (\ref{mucond}) and (\ref{quasi}).
In other words, one can calculate $\gamma_{\rm eq}$ 
from the following formula,
\begin{equation}
\gamma_{\rm eq} = \frac{n_{\rm b}}{P} \,\, \frac{dP(n_{\rm b})}{dn_{\rm b}}.
\label{gameq}
\end{equation}

\noindent
($ii$) {\it Frozen adiabatic index $\gamma_{\rm fr}$}.

\noindent
In this case all reactions of particle mutual transformations
are frozen, that is, $y_i=n_i/n_{\rm b}={\rm constant}$ for any particle species $i$.
The quasineutrality condition (\ref{quasi}) is then automatically satisfied 
and $\gamma_{\rm fr}$ can be calculated from the formula
\begin{equation}
\gamma_{\rm fr} = \frac{n_{\rm b}}{P} \,\, 
\frac{ \partial P(n_{\rm b}, \, y_{\rm e}, \, y_{\mu}, \, y_{\rm n},\, y_{\rm p},\, y_{\Lambda}, \, \ldots)}{ \partial n_{\rm b}}.
\label{gamfr}
\end{equation}

\noindent
($iii$) {\it Partly frozen adiabatic index $\gamma_{\rm part \, fr}$}.

\noindent
In this case all the slow reactions due to weak interaction 
(in particular, those with leptons ${\rm e}$ and $\mu$) 
are frozen, which means that 
\begin{eqnarray}
y_{\rm e}={\rm constant},
\label{ee}\\
y_{\mu}={\rm constant}.
\label{mumu}
\end{eqnarray}

In contrast, the reactions due to strong interaction 
are so fast that the matter is always in equilibrium 
with respect to them.
Here are these fast reactions
\begin{eqnarray}
{\rm p}+\Xi^-  &\leftrightarrow&  \Lambda + \Lambda, \\
{\rm n}+\Xi^0  &\leftrightarrow& \Lambda + \Lambda, \\
{\rm p}+\Sigma^- &\leftrightarrow& {\rm n} + \Lambda, \\
{\rm n}+\Sigma^0 &\leftrightarrow& {\rm n} + \Lambda, \\
{\rm n}+\Sigma^+ &\leftrightarrow& {\rm p} + \Lambda, \\
\Sigma^+ +\Sigma^- &\leftrightarrow& \Lambda + \Lambda.
\end{eqnarray}
and the corresponding conditions of equilibrium 
\begin{eqnarray}
\mu_{\rm p}+\mu_{\Xi^-} &=& 2 \mu_\Lambda, 
\label{11}\\
\mu_{\rm n}+\mu_{\Xi^0} &=& 2\mu_\Lambda, 
\label{22}\\
\mu_{\rm p}+\mu_{\Sigma^-} &=&\mu_{\rm n} + \mu_\Lambda, 
\label{33}\\
\mu_{\Sigma^0} &=& \mu_\Lambda, 
\label{44}\\
\mu_{\rm n}+\mu_{\Sigma^+} &=& \mu_{\rm p} + \mu_\Lambda, 
\label{55}\\
\mu_{\Sigma^+} +\mu_{\Sigma^-} &=& 2\mu_\Lambda. 
\label{66}
\end{eqnarray}

In stable NSs only $\Lambda$, $\Xi^{-}$, 
and (for the model {\bf GM1$^\prime$B}) $\Xi^0$-hyperons can be present,
so only the first two conditions, (\ref{11}) and (\ref{22}), are relevant.  

The final two conditions that should be taken into account are the
conservation of electric charge (\ref{quasi}) 
and the strangeness fraction $y_{\rm S}=S/n_{\rm b}$,
\begin{equation}
y_{\rm S}={\rm constant},
\label{yS}
\end{equation}
where  $S = \sum_i s_i n_i$ is the strangeness number density 
and $s_i$ is the strangeness of particle species $i$.
The condition (\ref{yS}) 
follows from the observation that strangeness is conserved
in reactions (\ref{11})--(\ref{66}) (while other reactions are frozen).

The conditions (\ref{quasi}), (\ref{ee}), (\ref{mumu}), and (\ref{11})--(\ref{yS})
allow one to express the pressure as a function of only four variables
$n_{\rm b}$, $y_{\rm e}$, $y_{\mu}$, and $y_{\rm S}$,
and to present adiabatic index $\gamma_{\rm part \, fr}$ in the form
\begin{equation}
\gamma_{\rm part \, fr} = \frac{n_{\rm b}}{P} \,\, 
\frac{ \partial P(n_{\rm b}, \, y_{\rm e}, \, y_{\mu}, \, y_{\rm S})}{ \partial n_{\rm b}}.
\label{gammapartfr}
\end{equation}
%

\section{Description of on-line material}
\label{appendix:online}

The results of our numerical calculations are summarized in a number of files 
that can be found on the web:

{\noindent}
{\bf http://www.ioffe.ru/astro/NSG/heos/hyp.html}.
We briefly describe them here.

\vskip 2mm
\noindent
%
(1) Files {\bf GM1A.dat}, {\bf GM1$^\prime$B.dat}, and {\bf TM1C.dat}
contain 
data concerning
the pressure $P$, energy density $\rho$ (both in MeV fm$^{-3}$)
and particle number densities $n_i$ (in fm$^{-3}$) 
for {\it three} models 
{\bf GM1A}, {\bf GM1$^\prime$B}, and {\bf TM1C}, 
studied in this paper.
Each file consists of 13 columns 
for 13 parameters listed in 
Tab.\ \ref{table:file_eos}.

\begin{table}
\begin{center}
\begin{tabular}[t]{cccccccc}
\hline\hline
Column number &  1 & 2 & 3 & 4 & 5 & 6 & 7  \\   
\hline
Parameter  & $n_{\rm b}$  & $P$  &  $\rho$   &  $n_{\rm e}$  & $n_{\mu}$  & $n_{\rm n}$  & $n_{\rm p}$ \\ \hline    
 Dimension \, & fm$^{-3}$ \, & MeV  fm$^{-3}$ \,  &  MeV  fm$^{-3}$  \, &  fm$^{-3}$  & fm$^{-3}$  & fm$^{-3}$  & fm$^{-3}$ \\
 \hline\hline\hline
 Column number &  8 & 9 & 10 & 11 & 12 & 13 &   \\   
 \hline
 Parameter  & $n_{\Lambda}$ & $n_{\Xi^-}$ & $n_{\Sigma^-}$ & $n_{\Xi^0}$ & $n_{\Sigma^0}$ & $n_{\Sigma^+}$ & \\
 \hline
 Dimension \, & fm$^{-3}$ &  fm$^{-3}$ & fm$^{-3}$ & fm$^{-3}$ & fm$^{-3}$ & fm$^{-3}$ & \\
 \hline\hline
\end{tabular}
\end{center}
\caption{Structure of the files {\bf GM1A.dat}, {\bf GM1$^\prime$B.dat}, and {\bf TM1C.dat}.}
\label{table:file_eos}
\end{table}

\vskip 2mm
\noindent
%
(2) Files  {\bf GM1A$\underline{\,\,\,\,}$Fields.dat}, 
{\bf GM1$^\prime$B$\underline{\,\,\,\,}$Fields.dat}, and {\bf TM1C$\underline{\,\,\,\,}$Fields.dat}
contain 
data concerning 
the values of meson fields (in MeV)
at different baryon number densities $n_{\rm b}$ (see Table \ref{table:Fields}).

\begin{table}
\begin{center}
\begin{tabular}[t]{ccccccc}
\hline\hline
Column number &  1 & 2 & 3 & 4 & 5 & 6 \\   
\hline
Parameter  & $n_{\rm b}$  & $\sigma$  &  $\omega^0$   &  $\rho_3^0$  & $\sigma^\ast$  & $\phi^0$ \\
\hline    
Dimension \, & fm$^{-3}$ \, & MeV &  MeV &  MeV & MeV  & MeV  \\
\hline\hline
\end{tabular}
\end{center}
\caption{Structure of the files  {\bf GM1A$\underline{\,\,\,\,}$Fields.dat}, 
{\bf GM1$^\prime$B$\underline{\,\,\,\,}$Fields.dat}, and {\bf TM1C$\underline{\,\,\,\,}$Fields.dat}}
\label{table:Fields}
\end{table}

\vskip 2mm
\noindent
%
(3) Files  {\bf GM1A$\underline{\,\,\,\,}$gamma.dat}, 
{\bf GM1$^\prime$B$\underline{\,\,\,\,}$gamma.dat}, and {\bf TM1C$\underline{\,\,\,\,}$gamma.dat}
contain 
data concerning 
the values of adiabatic indices $\gamma_{\rm eq}$, $\gamma_{\rm part\,fr}$, and $\gamma_{\rm fr}$ 
at different baryon number densities $n_{\rm b}$ (see Table \ref{table:adiabat}).

\begin{table}
\begin{center}
\begin{tabular}[t]{cccccccccc}
\hline\hline
Column number &  1 & 2 & 3 & 4 \\   
\hline
Parameter  & $n_{\rm b}$  & $\gamma_{\rm eq}$  &  $\gamma_{\rm part\,fr}$   & $\gamma_{\rm fr}$  \\
\hline    
 Dimension \, & fm$^{-3}$ \, & dimensionless &  dimensionless &  dimensionless   \\
 \hline\hline
\end{tabular}
\end{center}
\caption{Structure of the files {\bf GM1A$\underline{\,\,\,\,}$gamma.dat}, 
{\bf GM1$^\prime$B$\underline{\,\,\,\,}$gamma.dat}, and {\bf TM1C$\underline{\,\,\,\,}$gamma.dat}}
\label{table:adiabat}
\end{table}

\vskip 2mm
\noindent
%
(4) Files  {\bf GM1A$\underline{\,\,\,\,}$Mass.dat}, 
{\bf GM1$^\prime$B$\underline{\,\,\,\,}$Mass.dat}, and {\bf TM1C$\underline{\,\,\,\,}$Mass.dat}
contain 
data concerning 
the values of Landau effective masses $m^\ast_i$ [Eq.\ (\ref{meffLandau})]
at different baryon number densities $n_{\rm b}$ (see Table \ref{table:meff}).

\begin{table}
\begin{center}
\begin{tabular}[t]{cccccccccc}
\hline\hline
Column number &  1 & 2 & 3 & 4 & 5 & 6 & 7 & 8 & 9\\   
\hline
Parameter  & $n_{\rm b}$  & $m^\ast_{\rm n}$  &  $m^\ast_{\rm p}$   & $m^\ast_\Lambda$  & $m^\ast_{\Xi^-}$  & $m^\ast_{\Sigma^-}$ &
$m^\ast_{\Xi^0}$ & $m^\ast_{\Sigma^0}$ & $m^\ast_{\Sigma^+}$ \\
\hline    
 Dimension \, & fm$^{-3}$ \, & g &  g &  g & g  & g & g & g & g  \\
 \hline\hline
\end{tabular}
\end{center}
\caption{Structure of the files {\bf GM1A$\underline{\,\,\,\,}$Mass.dat}, 
{\bf GM1$^\prime$B$\underline{\,\,\,\,}$Mass.dat}, and {\bf TM1C$\underline{\,\,\,\,}$Mass.dat}}
\label{table:meff}
\end{table}

\vskip 2mm
\noindent
%
(5) Files  {\bf GM1A$\underline{\,\,\,\,}$F0.dat}, 
{\bf GM1$^\prime$B$\underline{\,\,\,\,}$F0.dat}, and {\bf TM1C$\underline{\,\,\,\,}$F0.dat}
contain dimensionless Landau parameters $F_0^{ij}$ (see Eq.\ \ref{eq:F.f}).
Note that $F_0^{ij}=F_0^{ji}$, so only $8 \times 9/2=36$ matrix elements
are independent and presented in these files (37 columns in each file; 
the first column is $n_{\rm b}$ in fm$^{-3}$).
An actual column number containing the Landau parameters with indices $i$ and $j$ 
can be found from Tab.\ \ref{table:F0}.
For instance, Landau parameters $F_0^{\Sigma^- \Xi^0}$ are given in the column 29.
Knowledge of the Landau parameters $F_0^{ij}$ and effective masses $m^\ast_i$
allows one to calculate the important thermodynamic derivatives, 
$\partial \mu_i(n_{\rm n}, \ldots,\, n_{\Sigma^+})/\partial n_j$.
As follows from Eqs.\ (\ref{eq:F.f}) and (\ref{dmudnj}),
\begin{equation}
\frac{\partial \mu_i(n_{\rm n}, \ldots,\, n_{\Sigma^+})}{\partial n_j}
=\frac{\partial \mu_j(n_{\rm n}, \ldots,\, n_{\Sigma^+})}{\partial n_i}
= \frac{{\mathrm \pi}^2 \, \hbar^3}{\sqrt{m^\ast_i  m^\ast_j \, p_{{\rm F}i}  p_{{\rm F} j}}}
\left( F_0^{ij} + \delta_{ij} \right).
\label{dmudnj2}
\end{equation}
%

\vskip 2mm
\noindent
%
(6) Files {\bf GM1A$\underline{\,\,\,\,}$F1.dat}, 
{\bf GM1$^\prime$B$\underline{\,\,\,\,}$F1.dat}, and {\bf TM1C$\underline{\,\,\,\,}$F1.dat}
contain dimensionless Landau parameters $F_1^{ij}$ and have exactly the same structure
as the files with $F_0^{ij}$ (see Tab.\ \ref{table:F0}).

\vskip 2mm
\noindent
%
(7) Files {\bf GM1A$\underline{\,\,\,\,}$Entr.dat}, 
{\bf GM1$^\prime$B$\underline{\,\,\,\,}$Entr.dat}, and {\bf TM1C$\underline{\,\,\,\,}$Entr.dat}
contain the symmetric entrainment matrix $Y_{ij}$ ($=Y_{ji}$). 
The first column is $n_{\rm b}$ in fm$^{-3}$; 
the next columns 2--37 are the elements of the entrainment
matrix $Y_{ij}$ 
[in cm$^{-3}$~erg$^{-1}$; see Eq. (\ref{eq:Y.F})] 
ordered in the same way 
as in the case of Landau parameters $F_0^{ij}$ and $F_1^{ij}$ 
(see Tab.\ \ref{table:F0}).

\begin{table}
\begin{center}
\begin{tabular}[t]{cccccccccccccc}
\hline\hline
Column number &  1 & 2 & 3 & 4 & 5 & 6 & 7 & 8 & 9 & 10 & 11 & 12 & 13 \\   
\hline
Parameter  & $n_{\rm b}$  & ${\rm nn}$  &  ${\rm np}$   &  ${\rm n} \Lambda$  & ${\rm n} \Xi^-$  & ${\rm n} \Sigma^-$  & ${\rm n} \Xi^0$ & ${\rm n} \Sigma^0$ & ${\rm n} \Sigma^+$ & 
${\rm pp}$ & ${\rm p} \Lambda$ & ${\rm p} \Xi^-$ & ${\rm p} \Sigma^-$ \\
\hline \hline \hline
Column number &  14 & 15 & 16 & 17 & 18 & 19 & 20 & 21 & 22 & 23 & 24 & 25 & 26 \\    
\hline 
Parameter  & ${\rm p} \Xi^0$  & ${\rm p} \Sigma^0$  &  ${\rm p} \Sigma^+$ &  $\Lambda \Lambda$  & $\Lambda \Xi^-$  & $\Lambda \Sigma^-$  & $\Lambda \Xi^0$ & $\Lambda \Sigma^0$ & $\Lambda \Sigma^+$ & 
$\Xi^- \Xi^-$ & $\Xi^- \Sigma^-$ & $\Xi^- \Xi^0$ & $\Xi^- \Sigma^0$ \\
\hline \hline \hline
Column number &  27 & 28 & 29 & 30 & 31 & 32 & 33 & 34 & 35 & 36 & 37 &  &  \\    
\hline
Parameter   & $\Xi^- \Sigma^+$  &  $\Sigma^- \Sigma^-$ &  $\Sigma^- \Xi^0$  & $\Sigma^- \Sigma^0$ & $\Sigma^- \Sigma^+$  & $\Xi^0 \Xi^0$ & $\Xi^0 \Sigma^0$ & $\Xi^0 \Sigma^+$ & 
$\Sigma^0 \Sigma^0$ & $\Sigma^0 \Sigma^+$ & $\Sigma^+ \Sigma^+$ &  & \\
 \hline\hline
\end{tabular}
\end{center}
\caption{A schematic structure of the files {\bf GM1A$\underline{\,\,\,\,}$F0.dat}, 
{\bf GM1$^\prime$B$\underline{\,\,\,\,}$F0.dat}, {\bf TM1C$\underline{\,\,\,\,}$F0.dat}, 
{\bf GM1A$\underline{\,\,\,\,}$F1.dat}, 
{\bf GM1$^\prime$B$\underline{\,\,\,\,}$F1.dat}, {\bf TM1C$\underline{\,\,\,\,}$F1.dat},
{\bf GM1A$\underline{\,\,\,\,}$Entr.dat}, 
{\bf GM1$^\prime$B$\underline{\,\,\,\,}$Entr.dat}, and {\bf TM1C$\underline{\,\,\,\,}$Entr.dat}.}
\label{table:F0}
\end{table}

\label{lastpage}

\end{document}